\documentclass{jpp}
\usepackage{graphicx}
\usepackage{marginnote}
\usepackage{setspace}

\usepackage{hyperref}
\usepackage[utf8]{inputenc}
\usepackage[T1]{fontenc}
\usepackage{amsmath}
\usepackage{colortbl}
\usepackage{enumitem}

\NewDocumentCommand{\vect}{>{\SplitList{,}}m}{%
  \begin{pmatrix}
    \ProcessList{#1}{\vectitem}
  \end{pmatrix}%
}
\newcommand{\pelon}{\mathring{\mathcal{E}}}

\newcommand{\xiv}{\sqrt{1-\lambda B}}

\NewDocumentCommand{\vectitem}{m}{#1\\}

\usepackage{xparse}
\ExplSyntaxOn
\NewDocumentCommand{\mymatrix}{m}
{
  \begin{pmatrix}
    \tl_set:Nn \l_tmpa_tl { #1 }
    \tl_replace_all:Nnn \l_tmpa_tl { ; } { \\ }
    \tl_replace_all:Nnn \l_tmpa_tl { , } { & }
    \tl_use:N \l_tmpa_tl
  \end{pmatrix}
}
\ExplSyntaxOff

\shorttitle{Near-axis measures of QI}
\shortauthor{E. Rodriguez, G. G. Plunk
}

\title{Near-axis measures of quasi-isodynamic configurations}

\author{E. Rodríguez\aff{1}, G. G. Plunk\aff{1}}

\affiliation{
\aff{1} Max Planck Institute for Plasma Physics, 17491 Greifswald, Germany
}

\begin{document}

\maketitle

\begin{abstract}
    We present a number of measures and techniques to characterise and effectively construct quasi-isodynamic stellarators within the near-axis framework, without the need to resort to the computation of global equilibria. These include measures of the reliability of the model (including aspect-ratio limits and the appearance of ripple wells), quantification of omnigeneity through $\epsilon_\mathrm{eff}$, measure and construction of MHD stabilised fields, and the sensitivity of the field to the pressure gradient. The paper presents, discusses and gives examples of all of these, for which expansions to second order are crucial. This opens the door to the exploration of how key underlying choices of the field design govern the interaction of desired properties (``trade-offs''), and provides a practical toolkit to perform efficient optimisation directly within the space of near-axis QI configurations.
\end{abstract}

\section{Introduction}
The range of stellarator magnetic confinement fields is broad \citep{spitzer1958stellarator,boozer1998stellarator,helander2014theory}, making it a daunting task both to understand and design them. This is true even when one restricts attention to particular subsets, known to have certain key properties such as \textit{omnigeneity} \citep{hall1975, bernardin1986,Cary1997,helander2014theory}.  Omnigenous fields are capable of confining, by definition, all collisionless charged particle orbits \citep{northrop1961guiding,littlejohn1983,wessonTok,blank2004guiding}, hence the particular interest in them. Even though the magnetic field magnitude $|\mathbf{B}|$ on their nested flux surfaces requires careful consideration \citep{boozer1983transport,nuhren1988,Cary1997,parra2015less}, there remains a wide range of choices to be made. This is true whether the problem is attacked from the optimisation perspective \citep{mynick2006} or a more controlled analytical approach. 
\par
One prominent case of the latter, which this paper shall be concerned with, is the \textit{near-axis description} of the field \citep{mercier1964equilibrium,Solovev1970,lortz1976equilibrium,garrenboozer1991a}. Consisting of an asymptotic description of the equilibrium field in the distance (described by $r$) from its centre (called the magnetic axis), the fields reduce to being described by only a few functions and parameters. These constitute a model for the field that has proved powerful in advancing the theoretical understanding of optimised omnigeneous stellarators \citep{mercier1964equilibrium,lortz1976equilibrium,landreman2020magnetic,landreman2021a,jorge2020naeturb,rodriguez2022phases,rodriguez2024maximum} as well as providing a practical tool in stellarator design \citep{landreman2019,landreman2022mapping,rodriguez2023constructing,jorge2022c,camacho-mata-2022}. The latter has however been restricted for the most part to a particularly mature subclass of omnigeneous stellarators: namely \textit{quasisymmetric} stellarators \citep{boozer1983transport,nuhren1988,rodriguez2020necessary,burby2020some}. Only recently \citep{plunk2019direct,camacho-mata-2022,jorge2022c,Camacho2023helicity,rodriguez2023higher,rodriguez2024near} has the near-axis framework of \textit{quasi-isodynamic} (QI) fields \citep{Cary1997,Helander_2009,Nührenberg_2010}, the other big class of optimised stellarators, been sufficiently developed. The lateness of this development has been a consequence of the complex treatment that these fields demand; while QS fields possess a direction of symmetry in $|\mathbf{B}|$ (either toroidal or helical), the QI fields do not (although they have closed poloidal contours). 
\par
The objective of this work is to help enable full use of the recent advances in the near-axis description of QI stellarators.  In particular, although \cite{rodriguez2024near} verified the construction of approximately QI fields to second order, it left open the issue of how to go about obtaining `good' solutions, {\em i.e.} how to best select input parameters, how to evaluate candidate fields and how to navigate the extremely large design space that the near-axis construction opens up. To address this we develop a set of measures and techniques using second-order near axis expansion to incorporate notions of neoclassical transport, MHD stability and equilibrium sensitivity into the construction of QI stellarators. With these tools in hand, we are in a position to construct and more thoroughly evaluate stellarator design candidates, and conduct systematic studies exploring things like optimization trade-offs.
\par
The paper is organized as follows. In Section~\ref{sec:defs} essential definitions and notation of near-axis theory are provided.  The rest of the paper is divided into sections describing  different types of measures.  In Section~\ref{sec:dB} we introduce the measure of field error $\delta\!B$ that quantifies the deviation of a finite aspect ratio consistent equilibrium from its asymptotic near-axis description, following \cite{landreman2021a}. This does however not qualify the physical impact of such deviations. Thus in section \ref{sec:eps_eff} we present calculations of the neoclassical $\epsilon_\mathrm{eff}$ effective ripple \citep{nemov1999} within the near-axis expansion, a form to quantify the lack of omnigeneity in which second order is central. In Section~\ref{sec:rw} we study the appearence of localised \textit{ripple} wells in the field, a feature that the near-axis description of $\epsilon_\mathrm{eff}$ cannot capture, focusing on finding what aspect ratio these first appear. Some questions of MHD stability are addressed in Section~\ref{sec:sens_shaping}, where we explore the sensitivity of the magnetic well \citep{greene1997,landreman2020magnetic} of a near-axis field to the choice of second order shaping. Finally, in Section~\ref{sec:shaf} we study the role played by the pressure gradient, and quantify the sensitivity of a given field to changes in plasma $\beta$. We conclude with some closing remarks.

\section{Definitions and notation}\label{sec:defs}

This paper does not re-derive the inverse-coordinate near-axis framework \citep{garrenboozer1991a,landreman2019}, but the work is best understood with some familiarity with how the equilibrium equations for an approximately QI field are solved through second order in the expansion. A full detailed and pedagogical account may be found in \cite{rodriguez2024near} and references within, but we provide some pointers on notation to help make the presentation as self-contained as possible. 
\par
In the vain of the near-axis expansion, all equilibrium quantities are expanded in a Taylor-Fourier series in powers of $r=\sqrt{2\psi/\bar{B}}$ (a pseudo-radial distance from the axis, where $\psi$ is $2\pi$ the toroidal magnetic flux and $\bar{B}$ a reference magnetic field) and cosine (sine) harmonic in the poloidal, $\theta$, (or helical, $\chi=\theta-N\varphi$ for $N\in\mathbb{Z}$) angle. Here the angles are part of the Boozer coordinate system \citep{boozer1981plasma}, which eases the description of the magnetic field. This way, any function $f$ is written as $f=\sum_{n=0}^\infty r^n f_n$ and $f_n=\sum_{m=0}^n (f_{nm}^c(\varphi)\cos m\chi+f_{nm}^s(\varphi)\sin m\chi)$, where the latter sum is over even or odd numbers depending on the parity of $n$.\footnote{We may use the following shorthand as well: $F_{11}^c=F_{1c}$, $F_{11}^s=F_{1s}$, $F_{22}^c=F_{2c}$, $F_{22}^s=F_{2s}$ for any function that $F$ may be.} This subscript notation \citep[Sec.~2.3]{rodriguez2024near} will be repeatedly used, and we refer to $n$ as the order (first, second, etc.) of the expansion.
\par
In the inverse-coordinate framework, a magnetic field is described by only a few functions. First, the magnetic field strength, $B=|\mathbf{B}|$, which in the near-axis notation reduces to a set of functions of $\varphi$, such as $B_{2c}$ at second order. Because of the inverse-coordinate nature of the expansion, we describe flux surfaces of the magnetic field through the functions $X,~Y$ and $Z$, which define the relative position of flux surfaces respect to the magnetic axis in the Frenet-Serret frame (normal, binormal and tangent respectively) of the latter \citep[Eq.~(2.2)]{rodriguez2024near}. Finally, the covariant components of the field, $\{G,I,B_\psi\}$, are also explicitly involved \citep[Eq.~(2.1)]{rodriguez2024near}, as is the rotational transform of the field, $\iota$.
\par
The governing equilibrium equations impose important relations between the components of all of these quantities involved, and that is what solving the near-axis expansion is focused on. It is however important to emphasise the ingredients required to uniquely define a field in this framework. An approximately QI near-axis field is uniquely defined to first order by the set $\{\kappa(\ell),\tau(\ell),B_0(\varphi),\bar{d}(\varphi),\alpha_1(\varphi)\}$, where the first two are the curvature and torsion defining the shape of a magnetic axis (which must have vanishing curvature points \citep{plunk2019direct,Camacho2023helicity,rodriguez2024near}), the third is the magnetic field along the axis as a function of the toroidal Boozer angle, and the latter two define $B_1$ \citep[Eq.~(2.4)]{rodriguez2024near},
\begin{equation}
    B_1=\kappa(\varphi)B_0(\varphi)\bar{d}(\varphi)\cos\left[\chi-\alpha_1(\varphi)\right]. \label{eqn:X1}
\end{equation}
These first order choices also uniquely determine the elliptical shaping of flux surfaces around the magnetic axis.
\par
At second order, the key degrees of freedom (inputs) are $\{p_2,X_{2c}(\varphi),X_{2s}(\varphi)\}$, where the former parameter is the pressure gradient and the latter two are related to triangular shaping. For the detailed role played by each of these, we again refer the reader to \cite{rodriguez2024near}.

\section{Field truncation error $\delta\!B$} \label{sec:dB}
The construction of QI fields using the near-axis expansion is rather non-trivial already at first order in the expansion. Different choices of magnetic axis and the shaping of elliptical cross-sections require a careful balance. Unless actively sought, such fields will tend to present particularly elongated and twisted shapes \citep{camacho-mata-2022}.
\par
Even when suitable parameters are found, we must always remember that the field resulting from the near-axis description is approximate in nature. The study and design of a stellarator field requires, in practice, the full non-expanded equilibrium field. Thus, we generally need to construct a finite aspect ratio version of the near-axis field, which will show some deviations from the purely asymptotic description. Significant deviations can lead to the loss of those properties of the field carefully instilled in the near-axis field. This particularly concerns $|\mathbf{B}|$, the field strength, which controls guiding centre dynamics of particles, and \textit{omnigeneity}. When such deviations are limited, the near-axis description may be deemed \textit{good}. This measure of \textit{goodness} has been previously used \citep{camacho-mata-2022,Camacho2023helicity} to gauge the suitability of 1st order near-axis QI fields as interesting candidates for optimised stellarators, and is thus of some interest.
\par
To define this \textit{goodness} more precisely, let us introduce the field strengths  $B_\mathrm{global}$ and  $B_\mathrm{nae}^{(1)}$. Define $B_\mathrm{global}$ as the field strength of the global equilibrium at its outermost flux surface $\psi=\psi_a$ parametrised by $\theta$ and $\varphi$, angular poloidal and toroidal Boozer coordinates \citep{boozer1981plasma}. This global equilibrium is meant to be a finite aspect ratio representation of the near-axis field, and we follow the prescription of \cite{landreman2019}. That is, given a first order near-axis field, we calculate the shape of the elliptical flux surface at some radius $r=a$ (i.e., $a=\sqrt{2\psi_a/\bar{B}}$), and use it as the outer boundary of a global nested-flux surface, vacuum field. The solution then defines $\mathbf{B}_\mathrm{global}$.
\par
The comparison of this global field can then be made to the first order near-axis one, evaluated at the edge, $B_\mathrm{nae}^{(1)}=B_0+a^2B_1$. We define the \textit{field error} as,
\begin{equation}
    \delta\!B=\sqrt{\int(B_\mathrm{global}-B_\mathrm{nae}^{1\mathrm{st}})^2\mathrm{d}\varphi\,\mathrm{d}\theta}\Bigg/\sqrt{\int B_\mathrm{global}^2\mathrm{d}\varphi\,\mathrm{d}\theta}; \label{eqn:dB}
\end{equation}
the normalised root-mean-square difference. Large values imply significant differences between the realised field and the near-axis model, and thus the expectation of them behaving quite differently. 
\par
Gauging the \textit{goodness} of a NAE field as defined in Eq.~(\ref{eqn:dB}) requires us to solve a global equilibrium. Whether one uses \texttt{VMEC} \citep{hirshman1983}, \texttt{DESC} \citep{dudt2020desc}, \texttt{GVEC} \citep{hindenlang2019gvec} or any other nested flux surface solver, computing $\delta\!B$ is numerically costly as compared to the near-axis field, impeding in addition its theoretical investigation. Can we make sense of this goodness of the field from within the near-axis framework?

\subsection{Near-axis perspective on $\delta\!B$}
By definition, $\delta\!B$ measures the discrepancy between the full-$B$ and the truncation of an asymptotic form. That is, $\delta\!B$ is, in a sense, a measure of the truncation error of the asymptotic expansion. The truncation error of an asymptotic series \citep[Sec.~3.5]{bender2013advanced} can be gauged by the next order in the expansion,\footnote{Note that this is true by definition for sufficiently small $r$, the expansion variable. In any form, the next order bears information about the truncation error.} and thus the calculation to 2nd order could possibly be used to estimate $\delta\!B$. There is however an important caveat: in the near-axis expansion, moving from order $N$ to order $N+1$ in the expansion brings new degrees of freedom. From 1st to 2nd, we may shape the triangularity of surfaces in an infinite number of ways. 
\par
Amongst this myriad of possibilities, we should pick that which most closely corresponds to our particular scenario. This was dealt with in \cite{landreman2021a} and \cite{landreman2019}, which used the 1st order near-axis field of interest at $r=a$ to impose the shape of the outermost boundary of a global equilibrium, which they then asymptotically described. Formally, the description of the global equilibrium field $\tilde{\mathbf{B}}$ requires a double expansion in $r$ (a near-axis expansion) and $a$ (a large aspect ratio expansion to deal with the `global' nature of the field). Asymptotically computing this field, and evaluating its difference with $B_\mathrm{nae}^{(1)}(r=a)=B_0+a B_1$ can be shown to be \citep[Sec.~5.2]{landreman2021a},\footnote{We note that the construction detailed in \cite{landreman2021a}, in particular Eq.~(C30), is not correct. This is apparent from the lack of symmetry of the expression respect to the mid-point of a field period. The expression should be correctly constructed taking the $N$-fold symmetry into account, $ \bar{f}^{(2)}(\varphi) =\int_0^\varphi \hat{B}(\varphi')\mathrm{d}\varphi' + \left(\frac{1}{2}-\frac{N\varphi}{2\pi}\right)\int_0^{2\pi/N}\hat{B}(\varphi')\mathrm{d}\varphi'-\frac{N}{2\pi}\int_0^{2\pi/N}\mathrm{d}\varphi''\int_0^{\varphi''}\hat{B}(\varphi')\mathrm{d}\varphi'$.}
\begin{equation}
    \tilde{B}_2 = a^2\tilde{B}_0^{(2)} + r^2(\tilde{B}_{20}^{(0)} + \tilde{B}_{2c}^{(0)}\cos2\chi + \tilde{B}_{2s}^{(0)}\sin 2\chi).\label{eqn:B_tilde}
\end{equation}
The first is a \textit{mirror term}, Eq.~(C24) in \cite{landreman2021a}, arises from the finite aspect ratio construction (expansion in $a$), and is necessary to guarantee a constant toroidal flux $\psi_a=\bar{B}a^2/2$. As such, it is a perturbation on the near-axis input $B_0(\varphi)$, the magnetic field on axis that sets the leading trapping well structure. The $r$-dependent terms in Eq.~(\ref{eqn:B_tilde}) correspond to the 2nd order near-axis expansion of $\tilde{B}$, whose shaping is chosen to match the outermost surface. They may therefore be found in a form analogous to the regular second order construction, solving Eqs.~(A41)-(A42) in \cite{landreman2019} using Eqs.~(C9)-(C12) in \cite{landreman2021a} and then applying Eqs.~(A34)-(A36) in \cite{landreman2019}. 
\par
The field error $\delta\!B$ may therefore be estimated as,
\begin{equation}
    \delta\! B_\mathrm{ar}=a^2\sqrt{\frac{1}{2\pi}\int_0^{2\pi}\left[(\tilde{B}_{20}^{(0)}+\tilde{B}_0^{(2)})^2+\frac{(\tilde{B}_{2c}^{(0)})^2}{2}+\frac{(\tilde{B}_{2s}^{(0)})^2}{2}\right]\mathrm{d}\varphi}, \label{eqn:dB_ar}
\end{equation}
where all fields have been evaluated at the boundary $r=a$.

\subsection{Numerical implementation}
We have implemented the computation of $\delta\!B_\mathrm{ar}$ numerically in \texttt{pyQIC}\footnote{The code may be found in \texttt{https://github.com/SebereX/pyQIC.git}, a significantly different version of the original code in \cite{jorge2022c}, and the particular branch is specified in the Zenodo repository.}, following the original work of \cite{landreman2021a}, and introducing all the appropriate generalisations that apply to QI fields, especially the non-uniform $B_0$ and correct handling of half helicity axes \citep{Camacho2023helicity,rodriguez2024near}. In practice, the resources devoted to finding $\tilde{B}$ quantities are analogous to those employed when solving the 2nd order equations of a near-axis equilibrium. In fact, the very equations involved are quite similar, and more specific details may be found in the code itself. 
\par
\begin{figure}
    \centering
    \includegraphics[width=0.8\textwidth]{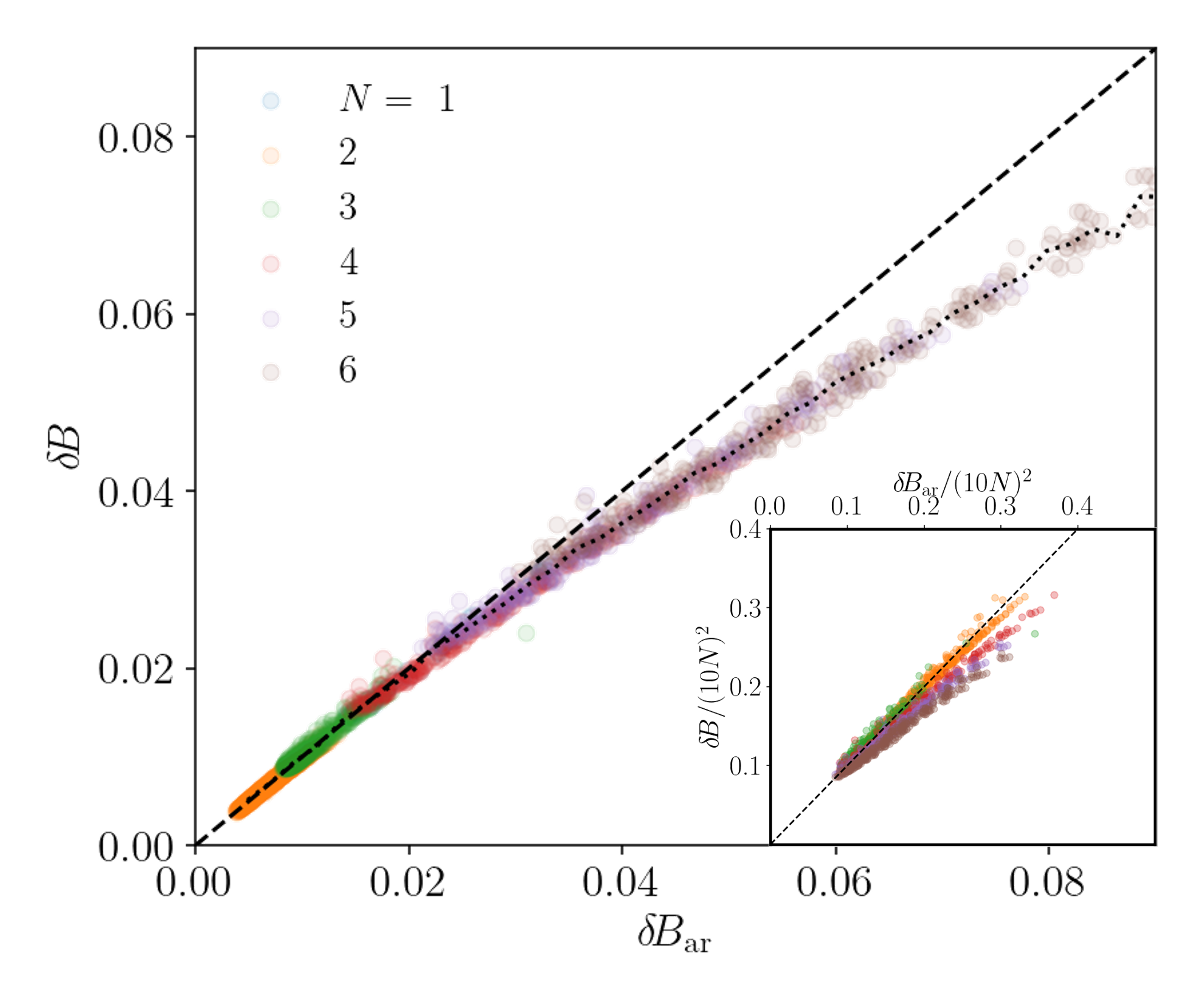}
    \caption{\textbf{Benchmark of the error field measure $\delta\!B_\mathrm{ar}$.} The main plot compares the field error $\delta\!B$, Eq.~(\ref{eqn:dB}), evaluated using the global equilibrium computed with \texttt{VMEC}, to the 
    near-axis estimate $\delta\!B_\mathrm{ar}$ for the QI near-axis configurations in the benchmark database of Appendix~\ref{app:QI_database}. The black broken line represents $\delta\!B=\delta\!B_\mathrm{nae}$, and the dotted one the moving average of the scatter. The colours for $\delta\!B_\mathrm{ar}$ denote the number of field periods of the configurations (see legend). The inset plot shows the same data scaled by the field period number as $1/N^2$. }
    \label{fig:delta_B_comparison}
\end{figure}

To benchmark $\delta\!B$ we need a set of near-axis configurations and their associated finite aspect ratio equilibria. The latter are obtained running the global flux-surface equilibrium code \texttt{VMEC} \citep{hirshman1983}, constructed using the near-axis surface at $r/R=0.07$ (roughly an aspect ratio of $A=14$). As for the benchmark set, we use a set of 1680 approximately QI, half-helicity near-axis fields (see some details in Appendix~\ref{app:bench} and a fuller description in an upcoming publication). To compute $\delta\!B$ as defined in Eq.~(\ref{eqn:dB}) the field strength is computed in Boozer coordinates using \texttt{BOOZXFORM} \citep{sanchez2000ballooning}. The comparison of $\delta\!B$ to the near-axis estimate $\delta\!B_\mathrm{ar}$ is shown in Figure~\ref{fig:delta_B_comparison}.
\par
The agreement is excellent (the correlation is in excess of 0.99), especially strong for the lowest $\delta\!B$ values. At larger $\delta\!B$, there is a clear systematic deviation, with the near-axis $\delta\!B_\mathrm{ar}$ overestimating the field discrepancy. This overestimation is a result of fields with larger $\delta\!B$ having a tendency for increased shaping and sensitivity, and thus smaller radii of convergence. Evaluation at a finite aspect ratio thus tends to overestimate the magnitude of the field. However, $\delta\!B_\mathrm{ar}$ retains a monotonic relation with $\delta\!B$.
\par
\par
The behaviour of the field error is clearly seen to depend on the number of field periods. They appear to form separate clusters, with the larger field period numbers leading to larger deviations (in agreement with the arguments before, with the exception of the special case $N=1$). Rescaling $\delta\!B/N^2$ (see inset of Figure~\ref{fig:delta_B_comparison}) clusters all different fields together. This quadratic scaling follows the involvement of second derivatives in $\varphi$ in the construction of the 2nd order near-axis equilibrium expansion (see \cite{rodriguez2024near}) and approximate scaling symmetries in axis shapes as discussed in \cite{plunk2025-geometric}. 
\par
All together, the error field near-axis measure $\delta\!B_\mathrm{ar}$ is an excellent predictor of $\delta\!B$. The systematic deviation may lead to a failure of a qualitative agreement at larger errors, but even this departure could be refined by using a fit of the curve in Figure~\ref{fig:delta_B_comparison} to map $\delta\!B_\mathrm{ar}$ to $\delta\!B$.

\section{Effective ripple: $\epsilon_\mathrm{eff}$} \label{sec:eps_eff}
Although the simplicity of the field error $\delta\!B$ makes it appealing as a measure of how `good' a given 1st order near-axis field is, it ignores the fact that not all field deviations (even if they may correspond to the same $\delta\! B$) affect the behaviour of the field equally. In particular, some deviations will spoil \textit{omnigeneity} more than others. The field error measure overlooks a second important point, namely that deviations from omnigenity are unavoidable to a certain degree \citep{Cary1997} even at first order \citep{plunk2019direct,rodriguez2023higher}, so it is important to separately evaluate such deviations at each order.  Thus, we are in need of some way of measuring the departure from ideal omnigeneity, of which there are many \citep{mynick2006,nemov2008poloidal,rodriguez2022measures,goodman2023constructing,dudt2024magnetic}.
\par
In this paper we consider the effective ripple $\epsilon_\mathrm{eff}$ as the measure of deviations from omnegeneity or ``omnigeneity error''. This is a geometric, dimensionless scalar quantity defined by \cite{nemov1999evaluation} whose magnitude (or more precisely, that of $\epsilon_\mathrm{eff}^{3/2}$) controls the neoclassical electron heat transport across a flux surface in a plasma that balances the radial losses with collisions; \textit{i.e.} the so-called $1/\nu$ regime. Given its clear physical interpretation, the evaluation of $\epsilon_\mathrm{eff}$ is commonly used to assess modern optimised stellarators, which aim at values below a percent. Here, we shall be interested in evaluating it within the near-axis framework. Doing so will: (i) avoid having to solve global equilibria and perform neoclassical calculations to assess fields, and (ii) provide insight on how near-axis choices affect it.
\par
Following \cite[Eq.~(29)]{nemov1999evaluation}, we express $\epsilon_\mathrm{eff}$ in our own notation as
\begin{equation}
    \epsilon_\mathrm{eff}^{3/2}=\frac{\pi (\bar{R}\bar{B})^2}{8\sqrt{2}}\lim_{L\rightarrow\infty}F(L)\,\int_{1/B_\mathrm{max}}^{1/B_\mathrm{min}}\mathrm{d}\lambda\, \lambda\sum_{j=1}^{N_\mathrm{well}}\mathcal{E}_j(\lambda), \label{eqn:def_eps_eff}
\end{equation}
where,
\begin{subequations}
    \begin{align}
    \mathcal{E}_j(\lambda) &=\frac{H_j(\lambda)^2}{I_j(\lambda)}, \label{eqn:Ej} \\
    H_j(\lambda) &= \frac{1}{\bar{B}^2}\int_{\ell_{j,L}}^{\ell_{j,R}}\mathcal{H}(\lambda,B)\frac{\mathbf{B}\times\nabla B\cdot\nabla \psi}{B} \mathrm{d}\ell, \label{eqn:Hj} \\
    I_j(\lambda) &= \int_{\ell_{j,L}}^{\ell_{j,R}}\frac{\xiv}{B}\mathrm{d}\ell, \label{eqn:Ij} \\
    F(L) &= \left(\int_0^L\frac{\mathrm{d}\ell}{B}\right)\left(\int_0^L|\nabla\psi|\frac{\mathrm{d}\ell}{B}\right)^{-2},
\end{align}
\end{subequations}
and
\begin{equation}
    \mathcal{H}(\lambda,B)=\frac{\xiv}{(B/\bar{B})^2}\left(\frac{4}{\lambda B}-1\right). \label{eqn:H_tilde}
\end{equation}
The factors $\bar{R}$ and $\bar{B}$ are reference length (major radius) and magnetic field magnitudes that normalise $\epsilon_\mathrm{eff}$ to make it a dimensionless quantity. The effective ripple is then to be understood as a sum over all trapped particles, $\lambda$, in all wells, labelled $j$, along the field line, running from $0$ to $L$ (in the limit of $L\rightarrow\infty$). That is, on an irrational surface, over the whole flux surface. Note that this involves a sum over very different classes of trapped particles, including particles that travel over a fraction or many periods of the torus, which makes the calculation particularly hard. 
\par
Each of these classes of particles contributes to the total ripple through $\mathcal{E}_j(\lambda)\geq0$, defined in Eq.~(\ref{eqn:Ej}), which for the field to be omnigeneous, must vanish for all trapped particles. The non-omnigeneous drive comes through $H_j$, an average between bounce points $\ell_{j,R/L}$ of the off-surface magnetic drift, $\mathbf{v}_d\cdot\nabla\psi$, weighted by the parallel velocity $\propto \sqrt{1-\lambda B}$. The heat flux, $\Gamma$, depends quadratically on $H_j$. This dependence stems from having $\Gamma \sim \delta\!f ~\mathbf{v}_d\cdot\nabla\psi$, where the perturbed particle distribution $\delta\!f$ arises from a balance between the average drift and collisions. The competing mixing rate of collisions is related to the function $I_j$, which is formally similar to the second adiabatic invariant. Finally $F(L)$ is a geometric normalisation factor that plays the role of normalising the off-surface projection of the drift. 
\par

\subsection{Identification of the expansion} \label{sec:eps_eff_adj}
Many of the complexities involved in evaluating $\epsilon_\mathrm{eff}$, Eq.~(\ref{eqn:def_eps_eff}), disappear when its asymptotic near-axis form is considered. To see this, let us start by introducing the asymptotic form of the magnetic field strength, which defines trapped particle classes and determines the radial drift. Concerned with the description of a magnetic field that is approximately QI, we can write $B\approx B_0(\varphi)+rB_1(\chi,\varphi)+r^2B_2(\chi,\varphi)$, where \citep{rodriguez2024near}
\begin{subequations}
    \begin{align}
        B_1 &= -B_0(\varphi)d(\varphi)\sin\left[\alpha-\alpha_\mathrm{buf}(\varphi)\right], \label{eqn:B1} \\
        B_2 &= B_{20}(\varphi) + B_{2c}(\varphi)\cos 2\chi + B_{2s}(\varphi)\sin2\chi.
    \end{align} \label{eqn:nae_B}
\end{subequations}
To zeroth order in $r$, $B_0(\varphi)$ defines (by assumption) a single toroidal trapping well per field period. The higher order contributions in $r=\sqrt{2\psi/\bar{B}}$ then deform this underlying well to make distinct ones for different field lines, labelled by $\alpha=\theta-\iota\varphi$. Asymptotically, they do not introduce additional wells. The particular form of $B_1$ in Eq.~(\ref{eqn:B1}) corresponds to that of an omnigeneous field (when taking $d$ to be an odd function respect to $B_0$, and $d=0$ at the bottom and top of the trapping well), except where $\alpha_\mathrm{buf}\neq0$, i.e. the buffer regions. These are for the most part unavoidable near the edges of the toroidal domain \citep{Cary1997,plunk2019direct,rodriguez2023higher}. The second order field contribution is for now chosen simply to satisfy stellarator symmetry, so $B_{20}$ and $B_{2c}$ are even, and $B_{2s}$ odd.
\par
As a result of the trapping wells becoming labelled by $\alpha$, the infinite sum over trapping wells in the effective ripple, Eq.~(\ref{eqn:def_eps_eff}), becomes by Weyl's lemma of equidistribution \citep{weyl1916gleichverteilung}\citep[Lemma 2.2]{stein2011fourier} (assuming an irrational rotational transform),
\begin{equation}
    \sum_{j=1}^{N_\mathrm{well}}f_j\approx \frac{N_\mathrm{well}}{2\pi}\int_0^{2\pi}f(\alpha)\mathrm{d}\alpha,
\end{equation}
where $N_\mathrm{well}$ is the number of wells within a field line portion of length $L$. That is, the sum over wells simply corresponds to an average over $\alpha$. 
\par
Changing variables from $\ell$ to $\varphi$ (with the appropriate Boozer Jacobian $\mathcal{J}=(G+\iota I)/B^2$ \citep[Eq.~(6.6.8b)]{d2012flux}), and writing flux surface averages in terms of $(\alpha,\varphi)$,
\begin{equation}
    \epsilon_\mathrm{eff}^{3/2}=\frac{\pi}{8\sqrt{2}}\frac{(\bar{R}\bar{B})^2}{\mathcal{G}^2}\,\int_{1/B_\mathrm{max}}^{1/B_\mathrm{min}}\lambda \hat{\mathcal{E}}(\lambda)\,\mathrm{d}\lambda, \label{eqn:def_eps_eff_nae}
\end{equation}
where
\begin{equation}
    \hat{\mathcal{E}}(\lambda) = \frac{1}{\pi}\int_0^{2\pi}\frac{\hat{H}(\lambda,\alpha)^2}{\hat{I}(\lambda,\alpha)}\,\mathrm{d}\alpha, \label{eqn:E_nae}
\end{equation}
and all remaining definitions can be found explicitly at the beginning of Appendix~\ref{app:eps_eff}. The functions $\hat{H}$ and $\hat{I}$ (directly linked to the previous $H_j$ and $I_j$ of \cite{nemov1999}), Eqs.~(\ref{eqn:H_hat_nae})-(\ref{eqn:I_hat_nae}), have been defined to be dimensionless, and $\mathcal{G}^2$ (directly related to $F$) has dimensions of $[L]^2$, which matches the presence of the normalising $\bar{R}$.
\par
In this form, the asymptotic expansion in powers of $r$ can be carried out explicitly, and we do so in Appendix~\ref{app:eps_eff}. The details may be found there, and we emphasise the importance of carefully considering bounce integrals and $1/\sqrt{1-\lambda B}$ factors (as in \cite{rodriguez2023precession} and \cite{rodriguez2024maximum}). In the main text we summarise the resulting expressions in $\epsilon_\mathrm{eff}^{3/2}\approx \epsilon_\mathrm{eff}^{3/2,(0)}+r^2\epsilon_\mathrm{eff}^{3/2,(2)}+O(r^4)$, and focus on their significance. 

\subsection{Contribution from buffer regions: $\epsilon_\mathrm{eff}^{3/2,(0)}$}
The leading order contribution in the distance from the magnetic axis $r$ to $\epsilon_\mathrm{eff}^{3/2}$ is a constant offset that sets a lower bound to the effective ripple value in the limit of $r\rightarrow0$. The leading order asymptotic form can be written purely in terms of first (and 0th) order quantities (see the details in Appendix~\ref{app:eps_eff}), 
\begin{equation}
    \epsilon_\mathrm{eff}^{3/2,(0)}=\frac{\pi}{8\sqrt{2}}\frac{(\bar{R}\bar{B})^2}{(\mathcal{G}^{(1)})^2}\int_{1/B_0^{\mathrm{max}}}^{1/B_0^{\mathrm{min}}}\lambda\frac{(h^{(1)})^2}{I^{(0)}}\mathrm{d}\lambda, \label{eqn:eps_eff_0}
\end{equation}
where
\begin{subequations}
\begin{align}
    h^{(1)} &=\int_{\varphi_-}^{\varphi_+}\frac{B_0}{\bar{B}}\mathcal{H}(\lambda,B_0)d\sin\alpha_\mathrm{buf}\,\mathrm{d}\varphi, \label{eqn:h_1}\\
    I^{(0)} &= \int_{\varphi_-}^{\varphi_+}\frac{\sqrt{1-\lambda B_0}}{(B_0/\bar{B})^2}\mathrm{d}\varphi, \label{eqn:I0} \\
    (\mathcal{G}^{(1)})^2 &= 2\left(\frac{1}{2\pi}\int_0^{2\pi}\mathrm{d}\alpha\int_0^{2\pi/N}\mathrm{d}\varphi\frac{\Psi^{(1)}}{B_0}\right)^2\Bigg/\left(\int_0^{2\pi/N}\frac{\mathrm{d}\varphi}{B_0^2}\right)
\end{align}
\end{subequations}
where 
\begin{equation}
    \Psi^{(1)}=\left[(X_{1c}\sin\chi-X_{1s}\cos\chi)^2+(Y_{1c}\sin\chi-Y_{1s}\cos\chi)^2\right]^{1/2}. \label{eqn:G2_Psi1}
\end{equation}
The expression $\epsilon_\mathrm{eff}^{3/2,(0)}$, as can be seen by inspecting $h^{(1)}$, measures the impact that buffer regions have on omnigeneity. If such buffers where not present (which may only happen if the rotational transform matches the helicity of the axis, $\iota_0=M$ \citep{plunk2019direct,Camacho2023helicity,rodriguez2024near}), then $\alpha_\mathrm{buf}=0$ and $h^{(1)}=0$, implying $\epsilon_\mathrm{eff}\rightarrow 0$ as $r\rightarrow0$. In general, though, they will make a finite contribution, and the expression above allows us to gauge its magnitude. It is a measure of the level of non-omnigeneity of the first order construction. Only if this is sufficiently small is it reasonable to proceed to the next order in the expansion (2nd order) to evaluate the omnigeneity at that order \citep{rodriguez2023higher,rodriguez2024near}.
\par
The expression for $h^{(1)}$, Eq.~(\ref{eqn:h_1}), is then a local measure: trapped particles that do not venture into buffer regions near the trapping well tops, such as deeply trapped ones, will not contribute by any amount. And thus, the smaller the buffer, the smaller the fraction of particles contributing. Those trapped particles that do, contribute to $h^{(1)}$ only through their  radial magnetic drift (proportional to $d$) in the buffer region. By construction, and to abide by the condition of pseudosymmetry that avoids losses at turning points of the magnetic field, $d=0$ is chosen to vanish at both the bottom and the top of the trapping well. Hence, we expect $d\approx0$ near the trapping well tops and thus in the buffer regions, automatically limiting the magnitude of the effective ripple. To make this somewhat more quantitative, consider the buffer to have a toroidal extent $\delta$ about the well top. For a magnetic axis with a curvature $\kappa$ that has a zero of order $v$ (fixing the form of $d=\bar{d}\kappa$ \citep{plunk2019direct,rodriguez2023higher}), and a field $B_0$ with a first derivative of order $u-1$, following Eq.~(\ref{eqn:h_1}), we expect a scaling $h^{(1)}\propto \delta^{u/2}\times \delta^v \times \delta\sim\delta^{u/2+v+1}$. The first factor comes from the slowing down of the particle speed near the tops; the second, from the low radial drift linked to $d$ itself; and the latter the size of the buffer domain. The natural impulse to take $\delta\rightarrow0$ has, though, dire consequences on the shaping of the resulting field \citep{plunk2019direct,camacho-mata-2022}. One could alternatively increase the order of the power of $\delta$, for instance, by making the top of the trapping well flatter. Without the need to resort to any of these, the behaviour of $I^{(0)}$ helps reduce the buffer contribution even further. This second-adiabatic-invariant-like factor is largest for barely trapped particles, reducing the contribution of the buffer to the effective ripple. This physically corresponds to these particles being more strongly diffused by collisions in $\lambda$ space, to a large extent because they travel for a longer time, and thus their drift contribution being offset to some degree. All in all, the contribution of the buffer tends to be small (see Figure~\ref{fig:eps_eff_bench}), setting a negligible lower bound on $\epsilon_\mathrm{eff}$ in the limit of $r\rightarrow0$. 

\subsection{Omnigeneity to second order: $\epsilon_\mathrm{eff}^{3/2,(2)}$}
Given the rather benign contribution to the effective ripple from 1st order, it is important to assess the next order contribtion to $\epsilon_\mathrm{eff}$. The evaluation of the order $r^2$ correction to the effective ripple is presented in detail in Appendix~\ref{app:eps_eff}. The calculation is significantly more involved than 1st order, requiring the next \textit{two} order corrections of many terms and even involving 3rd order quantities formally. Many of those terms do not, in practice, contribute significantly, and it is a great convenience to retain only 2nd order contributions.\footnote{One could carry out an analysis similar to that in Section~\ref{sec:dB}, to find the third order components corresponding to the finite build of a second-order finite aspect ratio equilibrium construction, but we shall not do that here.} For this reason, we write a reduced, simplified form of these contributions (see the Appendix~\ref{app:eps_eff} for a more detailed account of this) that focuses on the next order contribution to $H$,
\begin{equation}
    \epsilon_\mathrm{eff}^{3/2,(2)}=\frac{\pi}{8\sqrt{2}}\frac{(\bar{R}\bar{B})^2}{(\mathcal{G}^{(1)})^2}\int_{1/B_0^{\mathrm{max}}}^{1/B_0^{\mathrm{min}}}\lambda\frac{(h^{(2)})^2}{I^{(0)}}\mathrm{d}\lambda, \label{eqn:eps_eff_nae_2}
\end{equation}
where
\begin{subequations}
\begin{align}
    h^{(2)} &= -2\int_{\varphi_-}^{\varphi_+}\mathcal{H}(\lambda,B_0)\frac{\Delta B_{2c}^\mathrm{QI}}{\bar{B}}\,\mathrm{d}\varphi, \label{eqn:h2} \\
    \Delta B_{2c}^\mathrm{QI} &= B_{2c}^\mathrm{QI}-\frac{1}{4}\partial_\varphi\left(\frac{B_0^2d^2}{B_0'}\cos2\alpha_\mathrm{buf}\right), \label{eqn:qi_2nd_order}
\end{align}
\end{subequations}
and $B_{2c}^\mathrm{QI}=-(B_{2c}\cos2\bar{\iota}\varphi+B_{2s}\sin2\bar{\iota}\varphi)$. The expression in Eq.~(\ref{eqn:qi_2nd_order}) is equivalent to the omnigeneity condition at second order when it vanishes, as derived explicitly in \cite[Eq.~(32c)]{rodriguez2023higher}. Thus, this approximation to the asymptotic $O(r^2)$ behaviour of the effective ripple is driven by the 2nd order non-omnigeneous behaviour, and will generally dominate $\epsilon_\mathrm{eff}^{3/2,(0)}$ at a finite aspect ratio. 
\par
Unlike to leading order, in this case the non-omnigeneous contribution is spread over the entirety of the trapping well, wherever $\Delta B_{2c}^\mathrm{QI}\neq0$, a local measure of radial drift imbalance. An imbalance at some point $\varphi$ will affect all trapped particles with $\lambda<1/B_0(\varphi)$ that pass over this point. Although they feel such deviation from omnigeneity, this is not to say that they will necessarily behave in a non-omnigeneous fashion, as the bounce-integral in Eq.~(\ref{eqn:h2}) can lead to partial cancellations of $\Delta B_{2c}^\mathrm{QI}$, which can have either sign. However, it remains true that if $\Delta B_{2c}^\mathrm{QI}\neq0$, then $\epsilon_\mathrm{eff}^{3/2,(2)}\neq0$. 
\par
The above then describes the significance of the omnigeneity condition of \cite{rodriguez2023higher}, which is a property of 2nd order fields. As such, different choices of 2nd order shaping (namely, triangularity and Shafranov shift) will affect the omnigeneous behaviour of the field, and their choice becomes key in constructing the stellarator field. This is not to say that the 1st order near-axis construction no longer is important, as it will remain to strongly affect the amount and form of 2nd order shaping needed to make a field more omnigeneous. A clear example of this is the radial drift involved in Eq.~(\ref{eqn:qi_2nd_order}), which by increasing axis curvature (and thus $d$ for a controlled elongation of flux surfaces) will generally require stronger shaping. 
\par
One may ask how can the shape at 2nd order be chosen to \textit{omnigenise} a field, and if it is an example of an optimisation problem. As shown in \cite{rodriguez2024near}, Eqs.~(3.6)-(3.7), there is however a unique, closed form choice of shaping that exactly achieves $\Delta B_{2c}^\mathrm{QI}=0$. This excludes, though, the neighborhoods (in $\varphi$) of flattening points of the magnetic axis, where the shaping necessary to achieve this ideal omnigeneous behaviour would diverge unless the first order construction was specially chosen.  Excluding regions around flattening points (see further discussion below), and applying this choice of shaping, we say we have ``omnigenised'' the construction. 

\par



\subsection{Numerical implementation and benchmark}

\begin{figure}
    \centering
    \includegraphics[width=\textwidth]{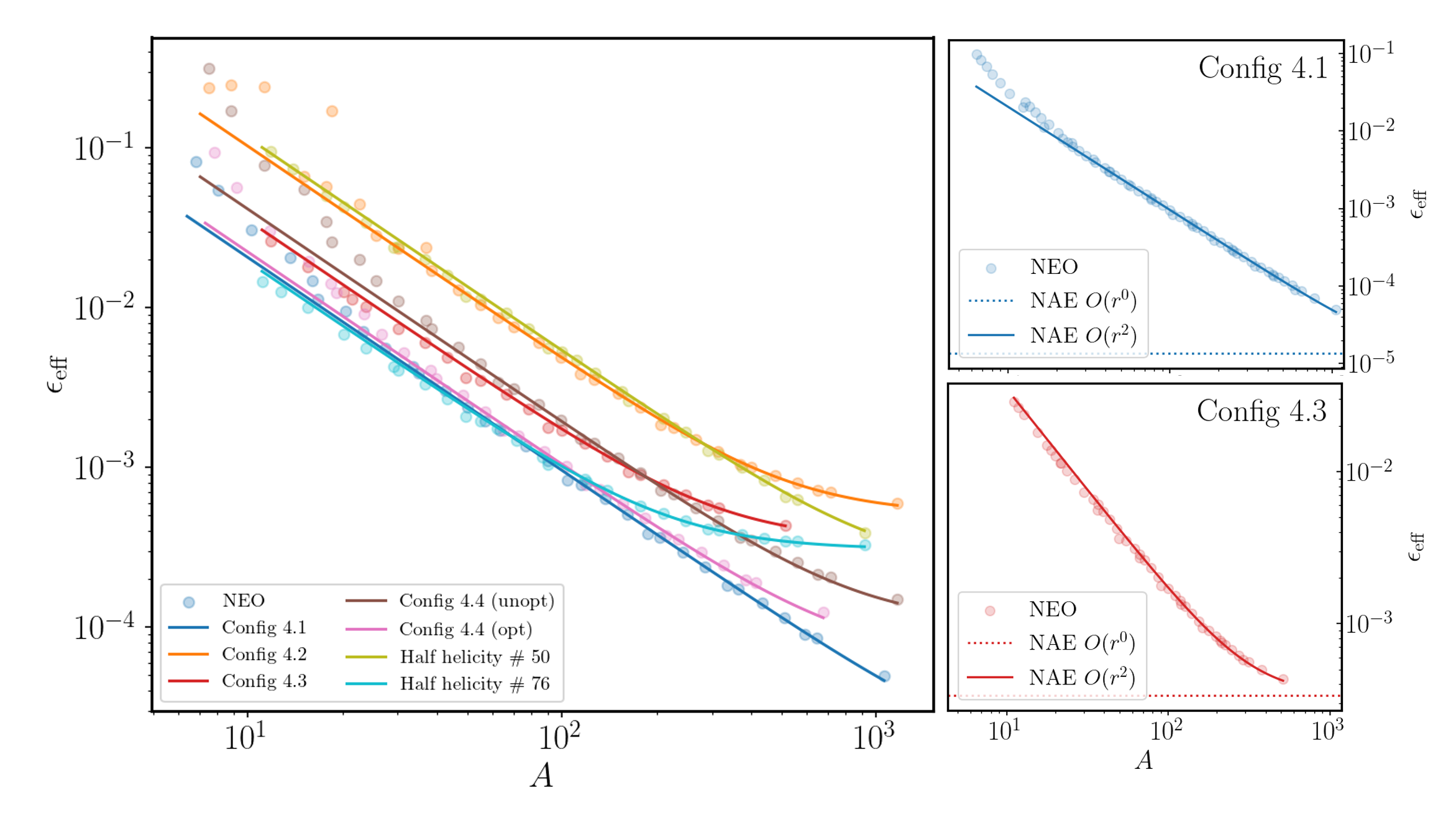}
    \caption{\textbf{Benchmark of effective ripple calculation.} The plots show, in log scale, a comparison of the effective ripple as calculated using the code \texttt{NEO} on global equilibria (scatter points) and calculated with the near-axis estimate (solid lines), for a number of different benchmark configurations (see Appendix~\ref{app:bench}). The two detail plots on the right show the individual comparison for two of the cases, including the 0th order ripple offset $\epsilon_\mathrm{eff}^{(0)}$ as reference (dotted line). The ripple is normalised to a reference $\bar{B}=1~$T and $\bar{R}=1~$m.}
    \label{fig:eps_eff_bench}
\end{figure}

Let us first benchmark of the near-axis estimate of the effective ripple above. To that end, we take an extended set of second-order near-axis equilibria previously used in \cite{rodriguez2024near}, described in Appendix~\ref{app:bench}, and their associated  global equilibria at a number of different aspect ratios. We then compare the near-axis estimate of $\epsilon_\mathrm{eff}$ (as a function of $r$) implemented in the \texttt{pyQIC} framework\footnote{The implementation makes use of the newer version of the \texttt{BAD} code for bounce averaging.}, to the effective ripple computed by the neoclassical code \texttt{NEO} of the finite aspect ratio equilibria. The comparison is presented in Figure~\ref{fig:eps_eff_bench} as a function of aspect ratio $A=R/r$.
\par
\begin{figure}
    \centering
    \includegraphics[width=0.7\linewidth]{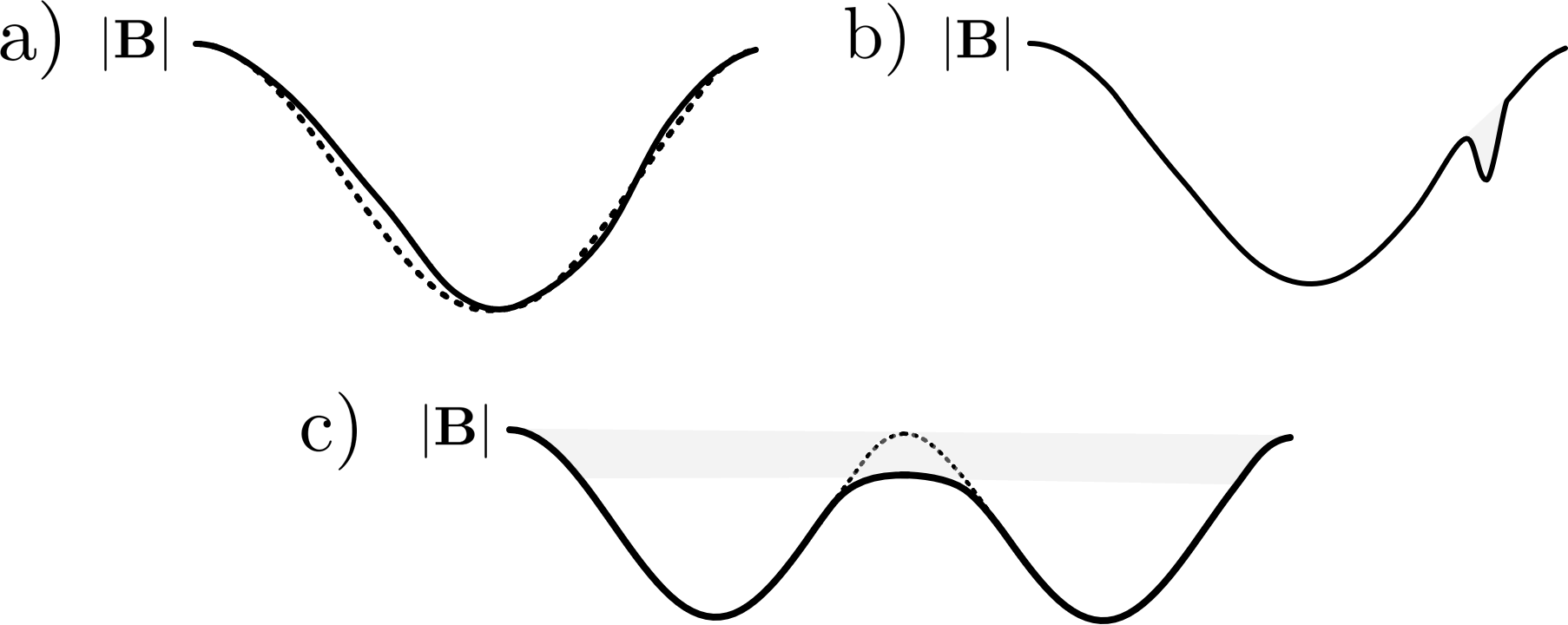}
    \caption{\textbf{Diagram with different contributions to $\epsilon_\mathrm{eff}$.} a) Deformation of $|\mathbf{B}|$ within the principal well violating the equal radial drift condition of omnigeneity (depicted by broken line). b) Appearance of local ripple or secondary wells. c) Misalignment of field maxima, leading to multiple-well trapped particles.}
    \label{fig:diag_deform_eps_eff}
\end{figure}

The agreement between the predicted near-axis $\epsilon_\mathrm{eff}$ and the finite volume equilibria calculation, although not exact, is excellent over quite a large range of $A$.  The buffer region contribution in all cases sets a rather benign lower bound to the value of $\epsilon_\mathrm{eff}$ at large aspect ratios, and seems to be in agreement with the global calculation. This shows the critical role played by the second order in the expansion, which controls the dominant $O(r^{4/3})$ behaviour.  At lower aspect ratios, $A\lesssim10$, the departures grow, as the deformation of the field increases and departs from its asymptotic description (see Figure~\ref{fig:diag_deform_eps_eff}). The latter is unable to capture local ripples \footnote{We do not consider the presence of $\theta$-independent sub-wells \citep{parra2015less}, and assume that any ripple will be localised in $\theta$, and thus will be non-omnigeneous.} and the appearance of new trapped particle classes (as in Figure~\ref{fig:diag_deform_eps_eff}b), nor misalignment of maxima (as in Figure~\ref{fig:diag_deform_eps_eff}c) with the associated abrupt modification of trapped particle behaviour. These effects become increasingly prominent at lower aspect ratios, and thus additional deviations between the near-axis and global calculation are to be expected.

\subsection{Measures of omnigeneity}
With the benchmark in place, we now introduce different measures based on the effective ripple to assess near-axis fields. We propose measures to characterise the effective ripple of 2nd order constructions, 1st order constructions, and the shaping involved when the latter are omnigenised at second order.
\par
We start with the fundamental form of the effective ripple,
\vspace{0.5cm}
\begin{enumerate}[label=(\alph*)]
    \item \quad\textit{Effective ripple from buffer regions}, $\epsilon_\mathrm{eff}^{3/2,(0)}$: \par
    \begin{quote} Leading order contribution to the effective ripple from the buffer regions, see Eq.~(\ref{eqn:eps_eff_0}). Its calculation requires information of 1st order, and the evaluation of bounce integrals between bounce points defined by $B_0(\varphi)$. \end{quote}
    \vspace{0.2cm}
    \item \quad \textit{Second order effective ripple}, $\epsilon_\mathrm{eff}^{3/2,(2)}$: \par
    \begin{quote} Approximate, order $O(r^{4/3})$ contribution to the effective ripple due to omnigeneity breaking at 2nd order, see Eq.~(\ref{eqn:eps_eff_nae_2}). It requires information of the 2nd order construction, and once again the evaluation of bounce integrals with bounce points defined by $B_0(\varphi)$.  \end{quote}
    \end{enumerate}
        \vspace{0.2cm}
    These combine into a single measure,
    \vspace{0.2cm}
    \begin{enumerate}[label=(\alph*), start = 3]
    \item \quad\textit{Effective ripple at the edge}, $\epsilon_\mathrm{eff}^\mathrm{edge}$: \par
    \begin{quote} Estimation of the effective ripple value at the flux surface corresponding to a reference aspect ratio of $A_\mathrm{ref} = 10$ evaluated using the asymptotic expressions 
    \begin{equation}
        \epsilon_\mathrm{eff}^\mathrm{edge}=\left(\epsilon_\mathrm{eff}^{3/2,(0)}+r_\mathrm{ref}^2\epsilon_\mathrm{eff}^{3/2,(2)}\right)^{2/3}.
    \end{equation}
    with $r_\mathrm{ref}=R/A_\mathrm{ref}$ and $R$ is the major radius calculated using the magnetic axis length, $L$, as $R=L/2\pi$. \end{quote}
\end{enumerate}
\vspace{0.2cm}
The $\epsilon_\mathrm{eff}^\mathrm{edge}$ is a dimensionless measure that can be directly interpreted as one would the effective ripple of an optimised stellarator. This is a measure that diagnoses the effective ripple of a fully consistent 2nd order near-axis construction. However, as discussed previously in this section and in \cite{rodriguez2024near}, there are certain aspects of the field that are a result of lower order choices in the near-axis construction. Thus, we would like to have some way to assess this ``intrinsic'' behaviour. The simplest measure is 
\vspace{0.2cm}
\begin{enumerate}[label=(\alph*), start = 4]
    \item \quad \textit{Non-omnigeneous mismatch at the well bottom}\footnote{We could define an equivalent measure at the top of the well. However, the behaviour at the minimum is of greater importance as all trapped particles experience it.}, $\Delta B_\mathrm{min}^\mathrm{QI}$:\par
    \begin{quote} Define,
    \begin{equation} 
    \Delta B_\mathrm{min}^\mathrm{QI}=r_\mathrm{ref}^2\left.\frac{\Delta B_{2c}^\mathrm{QI}}{\bar{B}}\right|_\mathrm{min}
    \end{equation}
    evaluated at the bottom of the trapping well. Here $\bar{B}$ is the average value of $B_0$ and $r_\mathrm{ref}$ a reference radial value (which we compute for $A_\mathrm{ref}=10$). The measure provides a sense of omnigeneity breaking near the bottom of the trapping well normalised to the average field strength. It only requires 1st order quantities, following Eq.~(3.8) in \cite{rodriguez2024near}. \end{quote}
\end{enumerate}
\vspace{0.2cm}
To translate this relative breaking of omnigeneity into the language of the effective ripple, we can evaluate $\epsilon_\mathrm{eff}^\mathrm{edge}$ for a near-axis field constructed using the omnigeneising shaping at 2nd order. That is, the shaping that forces $\Delta B_{2c}^\mathrm{QI}=0$ for all $\varphi$, barring some masked region near the flattening points. The shaping needed to achieve this is given in \cite{rodriguez2024near}, Eqs.~(3.6)-(3.7). The remaining ripple value is the result of the contribution from the straight sections, we denote
\vspace{0.2cm}
\begin{enumerate}[label=(\alph*), start = 5]
    \item \quad\textit{Effective ripple of omnigenised field}, $\epsilon_\mathrm{eff}^\mathrm{shap}$ \par
    \begin{quote} Estimate of the effective ripple of an omnigenised near-axis field at $r_\mathrm{ref}$. It is defined as $\epsilon_\mathrm{eff}^\mathrm{edge}$ for a field in which the 2nd order shaping has been chosen to enforce $\Delta B_{2c}^\mathrm{QI}=0$ everywhere except a 15\% of the toroidal domain around the inflection points. This evaluation requires solving the 2nd order near-axis equations with the appropriate omnigeneising shaping, and computing the appropriate bounce integrals. \end{quote}
\end{enumerate}
\vspace{0.2cm}

\par
To quantify how much shaping is required to construct this \textit{omnigenised} field, we define
\vspace{0.2cm}
\begin{enumerate}[label=(\alph*), start = 6]
    \item  \quad\textit{Average omnigenising shaping}, $\hat{T}^\mathrm{shap}$: 
    \begin{quote} 
    Averaged measure of the amount of shaping introduced at second order. We define 
    \begin{equation}
    \hat{T}^2=\frac{r_\mathrm{ref}^2}{2\pi}\int_0^{2\pi}\left(X_{20}^2+Y_{20}^2+Z_{20}^2+\frac{X_{2c}^2+X_{2s}^2+Y_{2c}^2+Y_{2s}^2+Z_{2c}^2+Z_{2s}^2}{2}\right)\,\mathrm{d}\varphi. \label{eqn:def_T_hat}
\end{equation}
    which we have non-dimensionalised with respect to the minor radius $r_\mathrm{ref}$, indicating the relative deformation of the elliptical cross-sections needed to omnigenise a field. We define $\hat{T}^\mathrm{shap}$ as the shaping value of the omnigenised field at an aspect ratio $A_\mathrm{ref}=10=R/r_\mathrm{ref}$.  Note that $\hat{T} = \hat{T}[X_{2c}, X_{2s}]$ may be considered a functional of the two free shaping functions at second order.
    \end{quote}
    \vspace{0.2cm}

The average measure $\hat{T}$ quantifies second order shaping in absolute terms, but makes no reference to the first order shape that it modifies. To elucidate how shaped a configuration is, it is important to consider the interaction with first order shaping at finite radius $r$.  To capture this, we introduce the critical radius $r_c$ (or its reciprocal, the critical aspect ratio $A_c\sim R/r_c$) defined in \cite{landreman2021a}. This measure captures the complexity of flux surfaces by reflecting the maximal radial extent at which flux surfaces continue to exist without any unphysical intersection, i.e. before the Jacobian of the coordinate system vanishes. That way, low $A_c$ indicates that the second order shaping is compatible with lower first order shaping for relatively compact fields. 
\vspace{0.2cm}
\item \quad\textit{Minimal aspect ratio of omnigenised field}, $A_c^\mathrm{shap}$: 
\begin{quote} Minimum value of the aspect ratio $A$ for which the near-axis construction of the omnigenised field does not present locally self-intersecting flux surfaces. This is based on the definition of \cite{landreman2021a},
\begin{equation}
    A_c=R/\mathrm{min}\left[r~|~\exists~\theta,\varphi: \mathcal{J}(r,\theta,\varphi)=0\right],
\end{equation}
where $\mathcal{J}$ is the Jacobian of the $\{r,\theta,\varphi\}$ coordinate system.

\end{quote}
\end{enumerate}
\vspace{0.5cm}

We gather the above measures evaluated on the configurations in the benchmark in Table~\ref{tab:eps_eff_bench}. The first figure to look at is the effective ripple at the edge, which provides a single measure to contextualise these near-axis constructions amongst the space of optimised stellarators. Configurations with values below a percent are often regarded adequate with regards to neoclassical behaviour. It is clear that the second order ($O(r^{4/3})$) contribution dominates the behaviour of $\epsilon_\mathrm{eff}$. To understand how much of the behaviour comes from the particular choice at 2nd order and how much is intrinsic to the lower orders, we then look at the properties of the omnigenised field.
\begin{table}
    \centering
    \begin{tabular}{c|c|c|c|c|c|c|c|}
       Configs &  4.1  & 4.2  & 4.3  & 4.4 (unopt)  & 4.4 (opt)  & \# 50  & \# 76  \\\hline
       $\epsilon_\mathrm{eff}^{3/2,(0)}\times 10^5$ & 0.01 & 1.75 & 0.63 & 0.16 & 0.08 & 0.33 & 0.52 \\
       $\epsilon_\mathrm{eff}^{3/2,(2)}$ [m$^{-2}$] & 0.290 & 3.079 & 0.643 & 0.790 & 0.313 & 3.829 & 0.263 \\
       $\epsilon_\mathrm{eff}^\mathrm{edge} [\%]$ & 2.46 & 13.5 & 3.46 & 5.47 & 3.10 & 11.4 & 1.91  \\
       \hline
        $\epsilon_\mathrm{eff}^\mathrm{shap} [\%]$ & 2.41 & 3.20 & 4.12 & 2.67 & 0.059 & 7.74 & 0.063  \\
       $\Delta B_\mathrm{min}^\mathrm{QI}$ [\%] & 5.78 & 0.94 & 2.53 & 0.94 & 0.08 & 4.17 & 0.21 \\       \hline\hline
      $ A_c $ & 2.09 & 5.84 & 2.38 & 4.46 & 4.93 & 3.00 & 1.59  \\\hline
       $A_{c}^\mathrm{shap}$ & 19.8 & 22.0 & 10.9 & 24.0 & 19.0 & 11.2 & 12.6 \\
       $\hat{T}^\mathrm{shap}$ & 2.27 & 0.78 & 0.30 & 0.75 & 0.79 & 0.24 & 0.30 
       
    \end{tabular}
    \caption{\textbf{Shaping configurations for effective ripple.} The table shows information regarding the omnigeneous nature of the near-axis constructions in the benchmark. The table is separated into to main parts. (Top) rows show the measures of omnigeneity, in particular the effective ripple for the original second order field in the benchmark (top), and the omnigenised form of the field (bottom). (Bottom) The lower rows present information regarding the shaping of the original field (top) and the omnigenised field. All the measures are defined in the main text. }
    \label{tab:eps_eff_bench}
\end{table}

The lack of omnigeneity near the inflection points is manifest in $\epsilon_\mathrm{eff}^\mathrm{shap}$, which does not approach the baseline value $\epsilon_\mathrm{eff}^{(0)}$ for any of the configurations. This is also indicated by the non-zero values of $\Delta B_\mathrm{min}^\mathrm{QI}$. However, the optimised version of config. 4.4 (by construction) and \#76, both show reduced ripple. What might appear most surprising, though, is that not all omnigenised configurations exhibit smaller ripple than their original forms, that is $\epsilon_\mathrm{eff}^\mathrm{shap}>\epsilon_\mathrm{eff}^\mathrm{edge}$ in some cases. This is a result of a strong non-omnigeneous drive near the bottom of the well, which in the original configuration partially cancelled the $\Delta B_{2c}^\mathrm{QI}$ away from it. 
\begin{figure}
    \centering
    \includegraphics[width=0.75\textwidth]{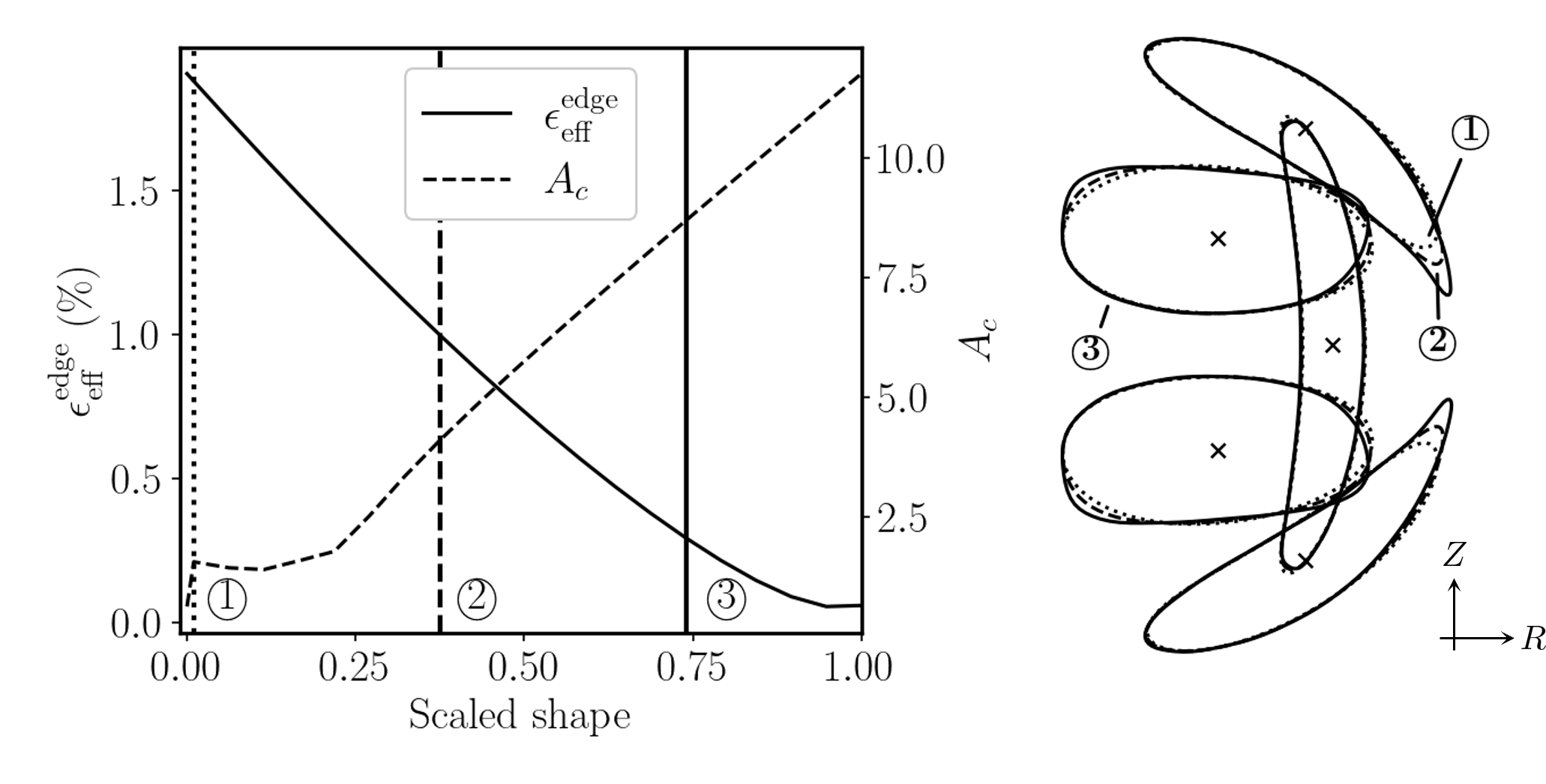}
    \caption{\textbf{Evolution of effective ripple with omnigeneising second order shaping.} The plots show the evolution of the half helicity benchmark configuration \#76 with changing 2nd order shaping. (Left) Evolution of $\epsilon_\mathrm{eff}^\mathrm{edge}$ (solid) and $A_c$ (broken) as a function of shaping. A value of 0 for the scaled shape corresponds to the original 2nd order benchmark configuration, while a value of 1 indicates the omnigenised construction. The shaping is scaled linearly in between. (Right) Examples of cross-section in cylindrical coordinates at three values of shaping, indicated as vertical lines on the left plot.}
    \label{fig:eps_eff_shaping}
\end{figure}
\par
Figure~\ref{fig:eps_eff_shaping} shows in more detail what the act of omnigeneising is, presenting the case of configuration \#~76. We choose this case because it has an intrinsic near-omnigeneous behaviour (small $\Delta B_\mathrm{min}^\mathrm{QI}$) at the bottom of its trapping well, and thus illustrates nicely the omnigeneising effect of shaping. We observe that moderate variations in the cross-sections lead to significant changes on the ripple. The plot on the left shows how by gradually increasing the shaping to its perfect omnigeneising value (0 meaning unshaped, and 1 the completely shaped case) reduces the ripple, while the changes in the surfaces lead to a significant increase in $A_c$. This reflects a general tendency observed with near-axis QI constructions: 2nd order shaping to improve omnigeneity is often costly in terms of shaping.  Freedom at first order must also be used to effectively construct omnigeneous fields.

\section{Appearance of secondary trapping wells: $A_w$} \label{sec:rw}
In the previous section we focused on assessing omnigeneity in terms of the asymptotic behavior of $\epsilon_\mathrm{eff}$. However, we noted that this did not account for the appearance of small localised secondary wells, which we now consider (see Figure~\ref{fig:diag_deform_eps_eff}b). The formation of these defects is linked to increased neoclassical transport \citep{ho1987} and particle losses \citep{mynick1983improved,paul2022energetic}. It is simple to picture a newly formed class of \textit{ripple} trapped particles, which live on those local ripples and thus experience whichever the local non-zero radial drift is.  
The deeper the well, the more trapped particles, and the further from bottom and top along the trapping well, the stronger the local drift. Having a sense of the presence of such ripples is then important.

\par

\subsection{Construction of a ripple well measure, $A_w$}
How do these secondary wells appear in the approximately QI near-axis scenario? By construction choice, in the limit of an infinite aspect ratio equilibrium ($r\rightarrow0$) there is a single well defined by $B_0$, and thus no trace of any secondary wells.
The ripple well question is then a finite aspect ratio one, which will require evaluating the asymptotic description at a finite $A$. We define $A_w$ as the largest aspect ratio $A$ for which a ripple well first appears along a magnetic field line within our asymptotic near-axis field model. Trapping wells are identified seeking points at which $\partial_\varphi|_\alpha B=0$ (i.e., partial toroidal derivative keeping the field line label constant, namely $\partial_\varphi|_\alpha B=(\partial_\varphi+\iota\partial_\theta)B$). To distinguish ripple wells from the main trapping well defined by $B_0(\varphi)$ on axis, we shall picture the appearance of secondary wells as a `dynamic' action in radius. As we look at surfaces of decreasing aspect ratio, ripples may start forming, but must do so by first forming inflection points, $\partial_\varphi|_\alpha^2 B=0$. We therefore define the ripple-well aspect ratio,
\begin{equation}
    A_w=R\big/\min\left\{r ~|~ \exists ~\theta,\varphi : \partial_\varphi|_\alpha B(r,\theta,\varphi)=0, \partial_\varphi^2|_\alpha B(r,\theta,\varphi)=0\right\}. \label{eqn:r_w}
\end{equation}
\par
In practice, finding $A_w$ requires a procedure (detailed in Appendix~\ref{app:rw}) similar to the calculation of the critical radius $r_c$ introduced above and originally in \cite{landreman2021a}. Briefly put, the two conditions in the definition of Eq.~(\ref{eqn:r_w}) written in terms of $B(r,\theta,\varphi)=\sum_{n=0}^2 r^nB_n(\theta,\varphi)$ may be interpreted as simultaneous algebraic equations on $r$ and $\theta$, which may be solved for every $\varphi$. The resulting multiple roots may be then written as a multi-valued function $\hat{A}_w(\alpha)$ (eliminating $\varphi$), of which the maximum is $A_w$.

\subsection{Assessment and application}
Let us consider the behaviour of ripple wells in two of the example configurations in the benchmark (see Appendix~\ref{app:bench}) and plot in Figure~\ref{fig:rrw_examples} the multi-valued function $1/\hat{A}_w$. For illustration purposes, we accompany these plots with the corresponding $B$ contours.

\begin{figure}
    \centering
    \includegraphics[width=\linewidth]{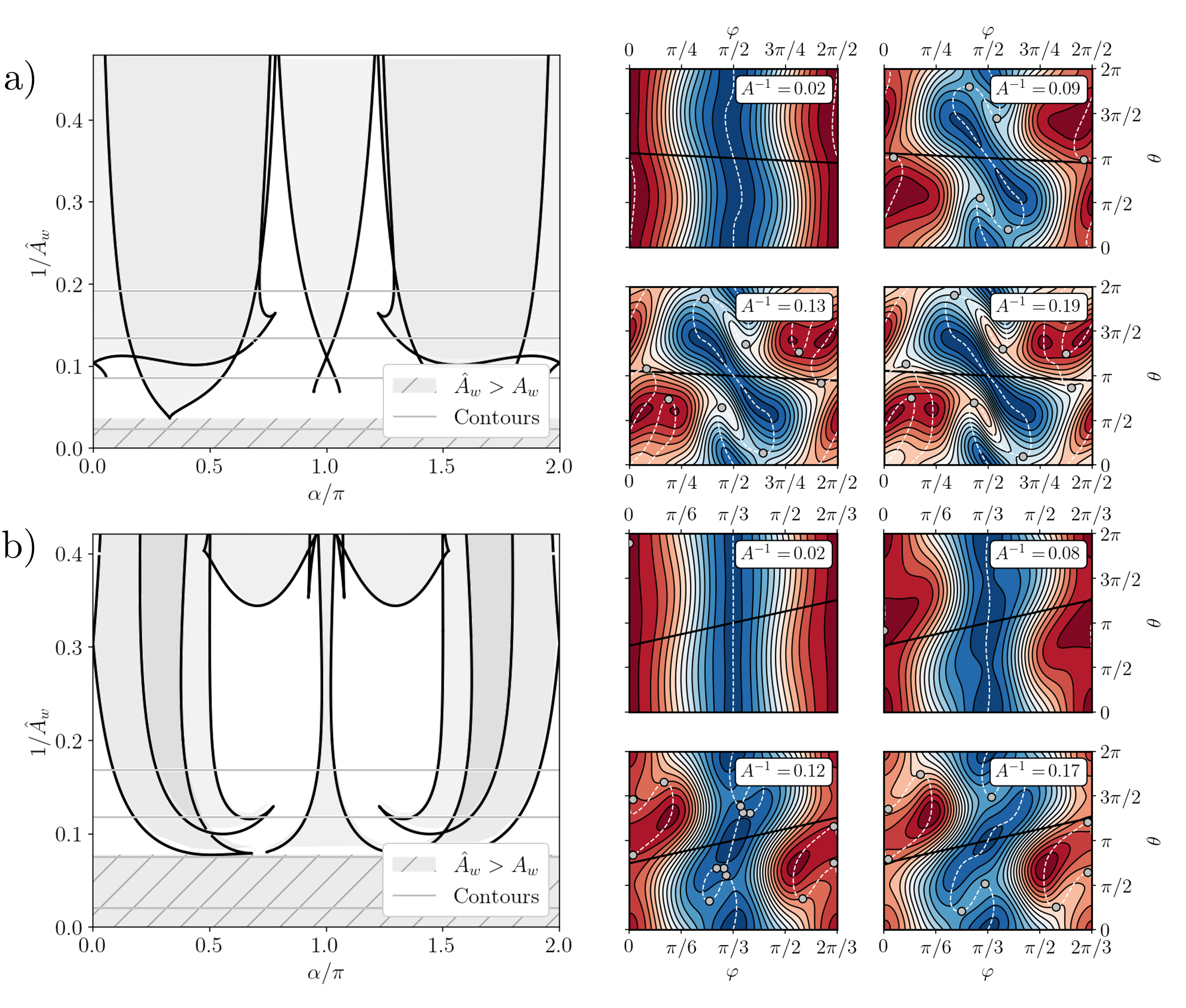}
    \caption{\textbf{Ripple well diagnostic measure $\hat{A}_w$ for configurations 4.1 and 4.3.} The plots show (left) the function $1/\hat{A}_w$ as a function of $\alpha$, the field line label, for Configs. 4.1. (a) and 4.3 (b), the \textit{spaghetti diagrams}. The hatched area represents $\hat{A}_w>A_w$, and the shaded regions the interval of $\alpha$ that have a ripple well. The right plots show the $|\mathbf{B}|$ contours corresponding to the $r$ values indicated by the horizontal silver lines on the left plot. The solid  black line in the contour plots shows the direction of a magnetic field line, the broken white line contours of $\partial_\varphi|_\alpha B=0$ and the scatter the corresponding inflections captured by the spaghetti diagram.}
    \label{fig:rrw_examples}
\end{figure}

The plots of $1/\hat{A}_w$, which we refer to as \textit{spaghetti diagrams} (borrowing from the condensed matter lingo), are best interpreted bottom to top. In the interval $0<1/\hat{A}_w<1/A_w$ the field does not present any secondary well, even though it may present topological defects in the contours of $|\mathbf{B}|$, i.e. \textit{puddles} \citep{rodriguez2023higher}. At $\hat{A}_w=A_w$, a \textit{band} is reached; that is, an inflection point appears along a field line for the first time. By symmetry, these inflections occur in pairs. Pairs of bifurcating branches appear and separate with increasing $1/\hat{A}_w$. The shaded areas these curves bound represent every field line (i.e., intervals of $\alpha$) containing secondary wells.  
This area grows at lower aspect ratios, and additional bifurcations do appear. We must mention the unpaired bifurcations that appear near $\alpha/\pi\approx 1$ of the top plot, which appear to contravene the general behaviour presented above, as they do not bound field lines with ripples. In fact, field lines in between have extended trapping wells. At least, until these branches cross. 
\par
With all this into account, at a finite $A<A_w$ we expect more ripples in config 4.1 as compared to 4.3.  What the spaghetti diagram fails to indicate is the location (in $\varphi$) of such ripples, and thus how strong the radial drift experienced by ripple trapped particles is. To fully assess the relevance of ripples, some of that information should be taken into account, but we shall not explore this further here, and leave the relationship between $A_w$ and the level of departure of the near-axis approximation to $\epsilon_\mathrm{eff}$ (see previous section) for future investigation.
\par

\begin{table}
    \centering
    \begin{tabular}{c|c|c|c|c|c|c|c|}
       Configs &  4.1  & 4.2  & 4.3  & 4.4 (unopt)  & 4.4 (opt)  & \# 50  & \# 76  \\\hline
       $ A_w$ & 27.4 & 37.0 & 12.9 & 32.8 & 24.4 & 18.5 & 16.5   \\
       $ A_c $ & 2.09 & 5.84 & 2.38 & 4.46 & 4.93 & 3.00 & 1.59  \\
    \end{tabular}
    \caption{\textbf{Ripple well distance in the benchmark configurations.} The table shows the values of the aspect ratio when the first ripple-well appears, $A_w$, and that at which the near-axis construction breaks down, $A_c$ for the configurations used as benchmark in the paper.}
    \label{tab:rw_bench}
\end{table}
The comparison of different configurations can be made more quantitative by looking at $A_w$, which we present in Table~\ref{tab:rw_bench}. The ripple measure $A_w$ exhibits some correlation with the shaping measure $A_c$ (they have a Pearson correlation of $\rho=0.75$, $p-\text{val}=0.05$). The more shaping, the more variation in the field, the larger the 2nd order field terms and thus the larger $A_w$. However, this is far from being a one-to-one relationship, and $A_w$ merits its own use as a field diagnostic. Minimising $A_w$ leads to `cleaner' $|\mathbf{B}|$ contours at finite aspect ratios. From this brief benchmark, it appears that half-helicity configurations (as well as the 2nd order QI optimised one) are more resilient to the appearance of wells. There are other elements in the near-axis equilibrium that have the potential to significantly influence the appearance of ripple wells, one being the shape of $B_0$. We would expect configurations with flatter minima, as those favouring MHD stability and maximum-$\mathcal{J}$ \citep{rodriguez2024maximum,plunk2024back}, to be more susceptible to the appearance of ripples. 
\par

\section{Magnetic well stability and shaping: $W$ and $\mathcal{G}^W$} \label{sec:sens_shaping}
There are crucial physical features of the field beyond omnigeneity that also depend on the second order form of the field. In particular, we are interested in MHD stability, which we capture in this paper through the so-called magnetic well criterion \citep{greene1997}: a field is deemed unstable (in particular, to interchange instability in the low plasma $\beta$ limit) if $V''<0$, where $V$ is the volume enclosed by flux surfaces and the primes denote derivatives with respect to $\psi$. In the context of the near-axis framework, $V''$ can be written in simple form as \citep[Eq.~(3.6)]{landreman2020magnetic},
\begin{equation}
    V''=2\pi\left|\frac{G_0}{\bar{B}}\right|\int_0^{2\pi}\frac{\mathrm{d}\varphi}{B_0^4}\left[3\left(B_{1s}^2+B_{1c}^2\right)-4B_0B_{20}-\frac{\mu_0p_2B_0^2}{\pi}\int_0^{2\pi}\frac{\mathrm{d}\varphi'}{B_0(\varphi')^2}\right]+O(r^2), \label{eqn:mag_well}
\end{equation}
where symbols have their standard near-axis meaning \citep{landreman2019,rodriguez2024near}. The key ingredient here is $B_{20}$, the poloidally averaged `radial' gradient of $|\mathbf{B}|$. A positive value is beneficial to stability as it tends to make field-lines curve outwards (from force balance, $\nabla_\perp(B^2/2)=B^2\pmb{\kappa} - \mu_0 \nabla p$ for $\pmb{\kappa}$ the curvature of field lines). The field component $B_{20}$ controls other physics such as precession, and the maximum-$\mathcal{J}$ property.  We do not delve into this here, but refer the reader to \cite{rodriguez2024maximum} for an in-depth discussion and several useful measures.
\par
Values of $V''$ are often reported as a \textit{relative well} measure, given in the form of a percentage \citep[Eq.~(14.29)]{Solovev1970}. Following this convention, we define the relative well of a finite aspect ratio equilibrium as $W=\int_0^{\psi_a}V''\mathrm{d}\psi/V'(0)$, where $\psi_a$ is the edge value of $\psi$. Such a measure might in general be misleading, as the magnetic well criterion is a locally radial one. In the present case of the near-axis expansion, where $V''$ is constant to leading order, though, $W$ is merely a  normalisation choice for $V''$. Choosing as a reference $r_\mathrm{ref}=R/A_\mathrm{ref}$ (unless otherwise stated, with $A_\mathrm{ref}\sim10$), in the near-axis framework, we define
\begin{equation}
    W=\frac{\bar{B}r_\mathrm{ref}^2}{4\pi G_0}\left[\int_0^{2\pi}\frac{\mathrm{d}\varphi}{B_0^2}\right]^{-1}V''.
\end{equation}
\par
With such a measure, we may assess the stability of a prescribed second order field. We expect the stability to depend both on shaping choices such as triangularity and Shafranov shift \citep[Eq.~(12.89)]{freidberg2014}, 2nd order quantities, but also first order ones. In fact, following the work in \cite{rodriguez2024maximum} and \cite{rodriguez2024near}, we can look near the straight sections of the field, where 2nd order shaping has no effect, to study the intrinsic stability contribution of the first order. In particular, the function $B_{20}$ will have some prescribed value there,
\vspace{0.2cm}
\begin{enumerate}[label=(\alph*)]
    \item \quad \textit{Relative radial gradient at the trapping well bottom}, $B_{20,\mathrm{min}}$:
    \begin{quote}
    We define $B_{20,\mathrm{min}}$ as
    \begin{equation} 
    B_{20,\mathrm{min}}=r_\mathrm{ref}^2\left.\frac{ B_{20}}{B_0}\right|_\mathrm{min}
    \end{equation}
    at the bottom of the trapping well. This is the poloidal-averaged radial gradient at the bottom of the well, which controls the poloidal precession of deeply trapped particles and contributes to stability if positive \citep{rodriguez2023precession, rodriguez2024maximum, plunk2024back}. This is an intrinsic feature of the first order construction. The measure is normalised to serve as a relative field strength change at a reference surface of aspect ratio $A_\mathrm{ref}$.  
    \end{quote}
\end{enumerate}
\vspace{0.2cm}

Away from these special points it is possible to choose the 2nd order shaping to make the field MHD stable, but how should this shape be chosen and how much does that shaping depend on the lower orders?

\par

\subsection{Sensitivity of vacuum well}
To determine the sensitivity of the magnetic well to the shaping at second order we compute the variation of the magnetic well defined in Eq.~(\ref{eqn:mag_well}) with respect to the input functions $X_{2c}$ and $X_{2s}$ that the integrand depends on. Following the general prescription detailed in Appendix~\ref{app:grad} for computing the shape gradient $\mathcal{G}^S=(\delta\!S/\delta\!X_{2c},~\delta\!S/\delta\!X_{2s})$ of a general functional of the form
\begin{equation}
    S=\int_0^{2\pi} f(B_{20},B_{2c},B_{2s},\varphi)\,\mathrm{d}\varphi,
\end{equation}
we may compute the shape gradient associated to the magnetic well, $\mathcal{G}^{W}$. For the magnetic well, we need, 
\begin{equation}
    \frac{\partial f}{\partial B_{20}}=-\frac{2r_\mathrm{ref}^2}{B_0^3}\left[\int_0^{2\pi}\frac{\mathrm{d}\varphi}{B_0^2}\right]^{-1}, \quad \frac{\partial f}{\partial B_{2c}}=0, \quad \frac{\partial f}{\partial B_{2s}}=0, \label{eqn:f_derivatives}
\end{equation}
which may be used to compute explicitly the gradient, as in Appendix~\ref{app:grad}.
\par
Although explicit, the resulting gradient involves the inversion of a differential linear operator (see Appendix~\ref{app:linear_system_2nd}). This is a result of $B_{20}$ being part of a self-consistent equilibrium solution. The necessary calculation to find the gradient is then computationally equivalent to solving a modified, second order near-axis equilibrium. The strength of this gradient calculation is that, because the magnetic well is linear on $B_{20}$, and thus also on the shaping, the gradient is independent of second order. Once we have computed the gradient, we know precisely how the magnetic well will change under changes in the shaping. 
\par
Before exploiting this knowledge of the gradient to concoct some diagnosing measures of the fields, we can gain some perspective on $\mathcal{G}^{W}$ by considering the simplest limit we can: that of an up-down symmetric tokamak field.\footnote{The considerations presented here hold beyond the context of fields with poloidally closed $|\mathbf{B}|$ contours. They extend to general magnetohydrostatic equilibrium near-axis field with the appropriate adjustments.} In this limit, any toroidal derivative drops out and (see details in the Appendix~\ref{app:grad_tok}), 
\begin{equation}
    \mathcal{G}_{2c}^W = \frac{3r_\mathrm{ref}^2}{\pi R}\frac{\bar{d}^4-1}{\bar{d}^4-3}, \quad \mathcal{G}_{2s}^{W} =0, \label{eqn:Vpp_tok}
\end{equation}
where $\bar{d}=1$ corresponds to a tokamak with circular cross-sections.\footnote{It is common to use $\bar{\eta}$ instead of $\bar{d}$ in the tokamak and quasisymmetric literature \citep{garrenboozer1991b,landreman2019}, but we here try to remain consistent with the rest of the notation in the paper.} For the case of circular cross sections, the magnetic well is insensitive to changes in shaping (triangularity) \citep{rodriguez2023mhd}. For a vertically elongated cross section ($\bar{d}<1$), an increase in $X_{2c}$ (i.e. of positive triangularity) favours the vacuum well, matching the well known tokamak intuition \citep{rodriguez2023mhd, freidberg2014}. The divergence as $\bar{d}^4\rightarrow 3$ is worrisome, as it leads to ill-behaviour for a very particular horizontally elongated elliptically shaped tokamak. Such a behaviour was previously described and analysed by \cite{rodriguez2023constructing} in the context of quasisymmetric stellarators, where the generalisation of the above is true. The divergence is indicative of only one \textit{particular form of shaping $X_{2c}$} being physical and consistent with how the construction is being carried out. This unphysicality can be seen on a similar divergence of triangularity, Eq.~(\ref{eqn:dd_dX2c}). Fortunately, configurations of interest tend to live away from this singularity. 
\par
Having a knowledge of the gradient $\mathcal{G}^{W}$ provides complete insight on the sensitivity of the magnetic well of a configuration. In particular, we are interested in knowing what is the minimal amount of shaping that would make a given 1st order near-axis configuration achieve a magnetic well (mirroring the practical approach in \citep{plunk2024back}). Because we have computed the shape gradient, we know precisely which shaping combinations will make a configuration stable: we simply need to choose $X_{2c}$ and $X_{2s}$ such that,
\begin{equation}
    \mathcal{C}=\int_0^{2\pi}\left(\mathcal{G}_{2c}^{W}X_{2c}+\mathcal{G}^{W}_{2s}X_{2s}\right)\,\mathrm{d}\varphi+W_\mathrm{ref}=0,
\end{equation}
where $W_\mathrm{ref}$ is the magnetic well value of the second order near-axis construction for $X_{2c}=0=X_{2s}$. This imposes marginal stability, and in the rare case of $W_\mathrm{ref}>0$, it would require relaxing the stability of the field. The `problem', though, is that there is no unique way of choosing the shaping to satisfy this constraint. We must then impose some regularising choice, which we take to be one that minimises shaping in the form of an integral over the sum of squares of all the second order shaping functions, as defined in Eq.~(\ref{eqn:def_T_hat}).\footnote{This is a natural regularising choice, $L^2$ norm, but others would also be possible (especially those that include penalties penalising large toroidal variations). Simpler forms could also be constructed only minimising the choice of input functions, which does however capture the notion of second order shaping less accurately.}
The problem of finding the minimal stabilising shaping can then be formulated as one of finding $X_{2c}$ and $X_{2s}$ that minimise the following constrained variational problem
\begin{equation}
    T^2[X_{2c},X_{2s}]=\hat{T}^2[X_{2c},X_{2s}]-\lambda~\mathcal{C}[X_{2c},X_{2s}], \label{eqn:T_variational}
\end{equation}
where $\lambda$ is a Lagrange multiplier. We denote the resulting shaping as $X_{2c}^\mathrm{mhd}$ and $X_{2s}^\mathrm{mhd}$, for which explicit expressions may be found in Appendix~\ref{app:variational_shape}. These expressions involve the shape gradients computed above, as well as some additional matrix multiplications, and could in principle be studied analytically. The main power of the approach is however providing a prescription to construct marginally stable approximately QI near-axis fields given a choice of axis, magnetic field strength on axis and first order information.
\par
To assess such \textit{stabilised} constructions, we introduce some simple measures. One of the simplest is,
\vspace{0.2cm}
\begin{enumerate}[label=(\alph*) ,start = 2]
    \item \quad\textit{Average second order shaping of stabilised field}, $\hat{T}^\mathrm{mhd}$:
    \begin{quote}
        The root mean square value of the consistent stabilised second order shaping, evaluated using Eq.~(\ref{eqn:def_T_hat}) using the minimal, stabilising choice of shaping, $\hat{T}^\mathrm{mhd} = \hat{T}[X_{2c}^\mathrm{mhd},X_{2s}^\mathrm{mhd}]$. It is therefore a simple measure of the amount of shaping necessary to stabilise a field, normalised to some reference $r_\mathrm{ref}$.
    \end{quote}
\end{enumerate}
\vspace{0.2cm}
A more acccurate measure of the shaping that takes toroidal variation into consideration may also be introduced, 
\vspace{0.2cm}
\begin{enumerate}[label=(\alph*), start = 3]
    \item \quad\textit{Critical aspect ratio of stabilised field}, $A_c^\mathrm{mhd}$:
    \begin{quote}
        The minimum aspect ratio for which the near-axis construction of the stabilised configuration is physical (non-intersecting flux-surfaces). That is $A_c$ for the near-axis field evaluated with $X_{2c}^\mathrm{mhd}$ and $X_{2s}^\mathrm{mhd}$. It is a dimensionless quantity.
    \end{quote}
\end{enumerate}
\vspace{0.2cm}
Finally, to get a sense of the sensitivity of the shaping and the magnetic well, we also define $\Delta A_c/\Delta W$,
\vspace{0.5cm}
\begin{enumerate}[label=(\alph*), start = 3]
    \item \quad\textit{Magnetic well sensitivity}, $ \Delta W/\Delta A_c$:
    \begin{quote}
        Change in the magnetic well $W$ with the change of shaping as measured by $A_c$ around the marginally stable point. Here we define it by finite differencing as,
        \begin{equation}
            \frac{\Delta W}{\Delta A_c} = \frac{W(V''=1~\text{T}^{-2}\text{m}^{-1})}{A_c(V''=1~\text{T}^{-2}\text{m}^{-1})-A_c(V''=0)},
        \end{equation}
        a dimensionless quantity. A larger value indicates increased sensitivity to shaping. We leave a more precise (perhaps analytic) form of this quantity to future work.
    \end{quote}
\end{enumerate}
\vspace{0.2cm}

\subsection{Numerical implementation and benchmark}

\begin{figure}
    \centering
    \includegraphics[width=\textwidth]{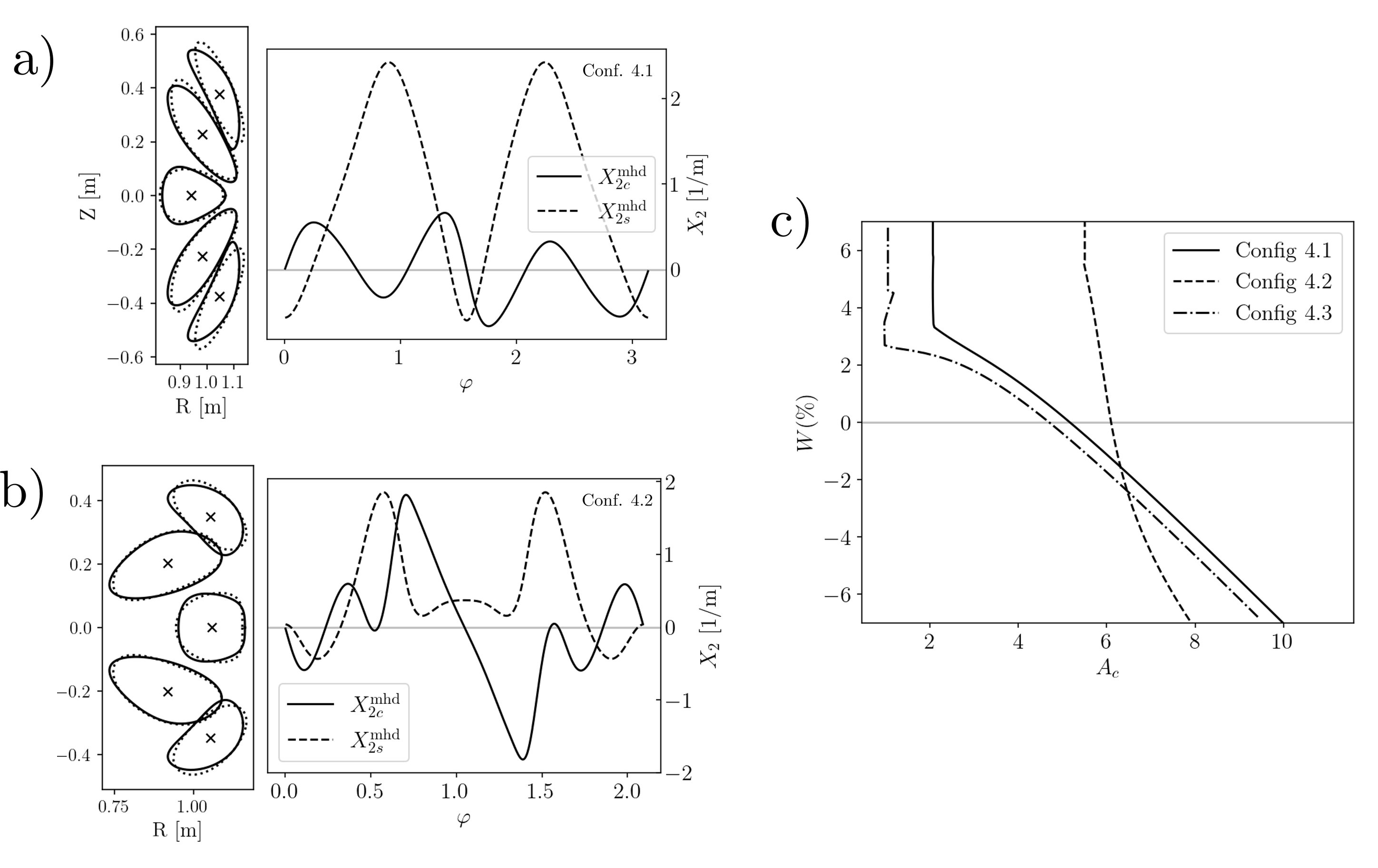}
    \caption{\textbf{Stabilised configurations and sensitivity of vacuum magnetic well to shaping.} (a)-(b) The plots show (left) the shaping input for stabilising the near-axis fields and (left) the resulting re-shaped cross-sections for the first three configurations of the benchmark. The black solid lines represent the re-shaped configuration, while the dotted one denotes the starting point. (c) Change in the magnetic well as a function of the shaping measure $A_c$, illustrating the sensitivity of the fields to shaping.}
    \label{fig:vac_well_examples}
\end{figure}

\begin{table}
    \centering
    \begin{tabular}{c|c|c|c|c|c|c|c|}
       Configs &  4.1  & 4.2  & 4.3  & 4.4 (unopt)  & 4.4 (opt)  & \# 50  & \# 76  \\\hline
       $B_{20,\mathrm{min}}$ [\%] & -2.15 & 0.80 & -2.10 & -1.37 & -0.99 & -5.27 & -0.52 \\
       $W$ [\%] & 5.78 & 5.56 & 4.52 & 9.12 & 8.52 & 6.64 & 4.12 \\
       $|W_\mathrm{mhd}| \times 10^{-16} $ & 1.53 & 1.39 & 2.10 & 1.78 & 2.97 & 5.00 & 0.48 \\\hline
       $A_c$ & 2.09 & 5.84 & 2.38 & 4.46 & 4.93 & 3.00 & 1.59 \\
       $A_c^\mathrm{mhd}$ & 5.17 & 6.10 & 4.70 & 6.43 & 6.23 & 5.91 & 4.82 \\
        $\hat{T}^\mathrm{mhd}$ & 0.20 & 0.17 & 0.16 & 0.22 & 0.19 & 0.22 & 0.13 \\
       $|\Delta W/\Delta A_c|$ [\%] & 1.30 & 8.48 & 1.25 & 2.43 & 3.29 & 1.37 & 1.20 \\
    \end{tabular}
    \caption{\textbf{Magnetic well sensitivity of benchmark configurations.} The table shows the values of the magnetic well and re-shaped fields for the benchmark configurations, the shape $\hat{T}$ and $r_c^\mathrm{mhd}$ second order shaping measures.}
    \label{tab:G_bench}
\end{table}

We apply the above measures to the configurations that constitute the benchmark for the this paper. We compute the shape gradient for all such configurations, compute their stabilised versions and evaluate the scalar measures presented above. A summary of the relevant vacuum well related properties are gathered in Table~\ref{tab:G_bench}. We start by noting that all configurations in the benchmark are MHD unstable, which appears to be the overwhelming tendency with constructed QI fields, as is the case with the other classes of omnigeneous fields \citep{landreman2020magnetic,landreman2022mapping,rodriguez2023constructing}). All examples encountered require some concerted shaping to push them towards MHD stability, which we illustrate in Figure~\ref{fig:vac_well_examples}. The correctness of the approach is evidenced in the vanishing of the magnetic well machine precision for the reshaped configurations, $W_\mathrm{mhd}$, in Table~\ref{tab:G_bench}. 
\par
Beyond the qualitative assessment of the shaping needed from the cross-sections, both the measure $\hat{T}$ and $A_c^\mathrm{mhd}$ quantitatively describe the amount of shaping, with the simpler form correlating with the latter. Whenever the re-shaped configuration has a small value for the critical radius, it means that significant shaping is present, and if there is a large difference from the original benchmark field construction, $A_c$ in Table~\ref{tab:G_bench}, we can conclude that the magnetic well is quite insensitive to shaping. Such sensitivity is more precisely measured by $\Delta W/\Delta A_c$. Note its large value in the case of config. 4.2, Figure~\ref{fig:vac_well_examples}b, where very minor changes in the shape if the cross-sections are enough to stabilise the field (see Figure~\ref{fig:vac_well_examples}c).  This configuration is also special in that $B_{20,\mathrm{min}}>0$, as we may expect from the help of having a finite plasma beta in that case. 
The high sensitivity is a potentially problematic feature of the field, if small changes in its shaping can change the response significantly. Note however that this sensitivity does change with $W$ (see Figure~\ref{fig:vac_well_examples}c), so that a given configuration could perhaps be made less sensisitive if sufficiently stabilised. This merits further study.
\par
If one was interested in finding the most MHD stable configurations, these measures could be used as a target of optimisation, which could be helped by exploitation of ideas from \cite{rodriguez2024maximum} and \cite{plunk2024back}. Before moving on, we note that this notion of \textit{MHD stability} is only a minimal one, but not necessarily sufficient to achieve true \textit{MHD stability}. Behaviour such as localised ballooning modes will react differently to the field geometry. In addition, this stabilising reshaping necessarily causes other 2nd order behaviour to change, and the ensuing tradeoffs are an important venue of future work.

\section{$\beta$ sensitivity of Shafranov shift } \label{sec:shaf}
So far in this paper we have made little explicit reference to plasma pressure. As we change the plasma pressure, though, we expect the magnetic field to change, as the Lorentz forces should continuously balance pressure gradients. One of the most notable consequences is the change in the Shafranov shift, the relative rigid displacement of nested flux surfaces with respect to each other. To describe this shift, we must define a reference point, a `centre', for each cross-section at different radii, and define the Shafranov shift to be the $\psi$-derivative of the position of that mid-point. In the context of the near-axis description, the Shafranov shift can be defined as \citep[Eqs.~(3.5)]{rodriguez2023mhd}\citep{landreman2021a},
\begin{equation}
    \pmb{\Delta} = \begin{pmatrix} \Delta_x \\ \Delta_y \end{pmatrix} =  \begin{pmatrix} X_{20}+X_{2c} \\ Y_{20}+Y_{2c} \end{pmatrix}, \label{eqn:shaf_shift_def}
\end{equation}
where the displacements are in the normal ($x$) or binormal ($y$) direction as defined by the signed Frenet-Serret frame of the magnetic axis. This expression matches the known definition in the tokamak limit, but one should bear in mind that there is no unique way in which to define the `centre' of a non-elliptical cross-section. 
\par
As the pressure gradient increases, so does the Shafranov shift, which results from the plasma pushing against the magnetic field (the poloidal field, and hence $\iota$, in a tokamak). Formally, as plasma pressure increases, so does $B_{20}$ linearly, and with it the rigid displacement of  surface $X_{20}$, eventually leading to two consecutive flux surfaces to touch (see Figure~\ref{fig:ellipses_shaf}). This break-down limits the existence of the equilibrium, and thus it is important to gauge how sensitive a field is to changes in pressure. In the context of the near-axis expansion the sensitivity of a field to changes in the pressure is most readily formulated keeping the shape of the magnetic axis fixed. This is unlike the more physical scenario of fixing some given coil currents (and thus a vacuum field), and letting plasma pressure grow. The latter requires separation of the vacuum from the plasma field, and performing a double expansion, and we do not do that here.
\par

\subsection{Shafranov shift shape gradient, $\mathcal{S}$}
The variations problem is now one of $\delta\!\pmb{\Delta}$, a change in the shift, with $\delta\!p_2$, a change in the pressure gradient. As a result of the change in the plasma pressure, from the definition in Eq.~(\ref{eqn:shaf_shift_def}), 
\begin{equation}
    \delta\!\pmb{\Delta}=\underbrace{\begin{pmatrix} \delta\!X_{20} \\ \delta\!Y_{20} \end{pmatrix}}_\text{Rigid shift} + \underbrace{\begin{pmatrix} 0 \\ \delta\!Y_{2c} \end{pmatrix}}_\text{Triangular shaping},
\end{equation}
where we distinguish between the shift of the `cross-section' due to a rigid displacement and due to triangular shaping. Although both contribute to the Shafranov shift, we shall here focus on the $\theta$-average, rigid shift, most useful when thinking of the problem in terms of shifted ellipses (see Figure~\ref{fig:ellipses_shaf}).
\par
The variation required then simply involves the 2nd order equations in Appendix~\ref{app:linear_system_2nd}, explicitly in Eq.~(\ref{eqn:LFf}). In this case, unlike for the magnetic well sensitivity, we need the variation respect to $\delta\!p_2$, a scalar,
\begin{equation}
    \hat{\mathbb{L}}\begin{pmatrix} \delta\!X_{20} \\ \delta\!Y_{20} \end{pmatrix} = \frac{\delta\!\pmb{f}}{\delta\!p_2}\delta\!p_2, \label{eqn:shaf_shape_eqn}
\end{equation}
the variation of $\hat{\mathbb{L}}$ vanishing exactly. Solving this linear system we may define, 
\begin{equation}
    \pmb{\mathcal{S}}=\hat{\mathbb{L}}^{-1}\frac{\delta\!\pmb{f}}{\delta\!p_2},
\end{equation}
which is a vector function of $\varphi$. This gradient holds for all 2nd order choices, as the equation is linear on $p_2$. That is, $\pmb{\mathcal{S}}$ is a property of the first order field. 
\par
It is natural to introduce some normalisation for the gradient now. With $[\Delta]=[L]^{-1}$ and $[p_2]=[p][L]^{-2}$, introducing the dimensionless plasma beta parameter $\beta_{p,2}=\mu_0p_2/(\bar{B}^2/2)$ for some reference magnetic field and the reference $r_\mathrm{ref}=R/A_\mathrm{ref}$, we define $\hat{\mathcal{\pmb{S}}}=(\bar{B}^2/2\mu_0 r_\mathrm{ref})\mathcal{\pmb{S}}$. We should therefore interpret this gradient as a fractional displacement of the surface at a reference aspect ratio $A_\mathrm{ref}$ due to a change in plasma beta.  We define
\vspace{0.2cm}
\begin{enumerate}[label = (\alph*)]
    \item \quad \textit{Maximum sensitivity of Shafranov shift}, $\hat{\mathcal{S}}_\mathrm{max}$:
    \begin{quote}
        Maximum relative value of the Shafranov shift gradient respect to changes of plasma $\beta$ at a reference aspect ratio $A_\mathrm{ref}$,
        \begin{equation}
            \hat{\mathcal{S}}_\mathrm{max}= \max_\varphi \left[\frac{\bar{B}^2}{2\mu_0 r_\mathrm{ref}}|\mathcal{\pmb{S}}|\right]. \label{eqn:S_max}
        \end{equation}
    \end{quote}
\end{enumerate}
\vspace{0.2cm}
\par
It is illuminating to consider the familiar tokamak limit. Simplifying the system (see details in the Appendix~\ref{app:2nd_order_simp_tok}), and considering for simplicity an up-down symmetric configuration, 
\begin{equation}
    \hat{\mathcal{S}}_\mathrm{max}=\frac{A_\mathrm{ref}}{\iota_0^2}\frac{\bar{d}^4}{\bar{d}^4-3},
\end{equation}
where all quantities have their usual meaning. The sensitivity grows with aspect ratio, as the same displacement becomes relatively larger compared to the minor radius. The scaling with $\propto\iota_0^{-2}$ responds to the physical picture of the poloidal field balancing the pressure gradient. The stronger that poloidal component, the more resilient to the push of the pressure. The role of elongation is more involved, although it guarantees for vertically elongated configurations, $\bar{d}<1$, $\hat{\mathcal{S}}_{x}<0$. That is, the there will be bunching of surfaces in the outboard side. It also shows the artificial divergence discussed in the previous section.
\par
So far, the gradient measure defined in Eq.~(\ref{eqn:S_max}), $\hat{S}_\mathrm{max}$, provides a scalar measure of sensitivity without distinguishing direction. The significance of any given shift does however depend on the direction in which it occurs. It is not the same to rigidly shift nested ellipses in the direction of the minor axis or the major axis. For directions in which surfaces are closer to each other, the bunching of surfaces is easier, and so, potentially, is their intersection.  We define an associated critical $\beta$ as that value for which, due to the rigid Shafranov shift, the underlying first order ellipses just touch at a radius of $r_\mathrm{ref}$. This definition directly uses $\hat{\pmb{\mathcal{S}}}$, as described in detail in Appendix~\ref{app:shaf}, and illustrated in Figure~\ref{fig:ellipses_shaf}.
\vspace{0.2cm}
\begin{enumerate}[label=(\alph*), start =2]
    \item \quad \textit{Estimate of critical plasma $\beta$}, $\beta_\Delta$:
    \begin{quote}
        Estimate of the critical plasma $\beta_p$ at which, due to the rigid part of the Shafranov shift, the first order near-axis elliptical flux surfaces of an equilibrium just touch at $r=r_\mathrm{ref}$. The value of the plasma beta is defined as a dimensionless scalar $\beta_\Delta$, for which a full definition is provided in Eqn.~\ref{eq:beta-Delta}. Large values of $\beta_\Delta$ indicate resilience of the Shafranov shift to changes in plasma $\beta$.
    \end{quote}
\end{enumerate}
\vspace{0.2cm}
The metric $\beta_\Delta$ condenses the information of the gradient ${\bf \mathcal{S}}$ into a single physically meaningful measure. However, it does so under certain simplifying assumptions. In particular, when it comes to describing when flux surfaces touch each other, it ignores the triangular shaping of flux surfaces, approximating them as ellipses. Thus $\beta_\Delta$ is only an estimate of the true critical plasma $\beta$, which we may compute numerically.
\vspace{0.2cm}
\begin{enumerate}[label=(\alph*), start =3]
    \item \quad \textit{Numerical critical plasma $\beta$}, $\beta_c$:
    \begin{quote}
        Critical value of $\beta_p=r_\mathrm{ref}^2\beta_{p,2}$, for an assumed aspect ratio $A_\mathrm{ref}$, at which the flux surfaces of the near-axis construction just touch.  That is, for $\beta_p =  \beta_c$ we have $A_c = A_\mathrm{ref}$.
    \end{quote}
\end{enumerate}
\vspace{0.2cm}

In other words, fixing $\beta_p = \beta_c$ the near-axis construction will not succeed for aspect ratios below $A_\mathrm{ref}$.  The value of $\beta_c$ may be found by a root search (and thus is more numerically costly) and includes all elements of shaping up to (and including) 2nd order.  We emphasise that this quantity, although defined in exact terms, may still not be the most realistic $\beta$ limit, because it is calculated taken the axis fixed, allowing flux surfaces to move around it as the plasma beta is changed. 

\subsection{Implementation and examples}
\begin{figure}
    \centering
    \includegraphics[width=0.8\textwidth]{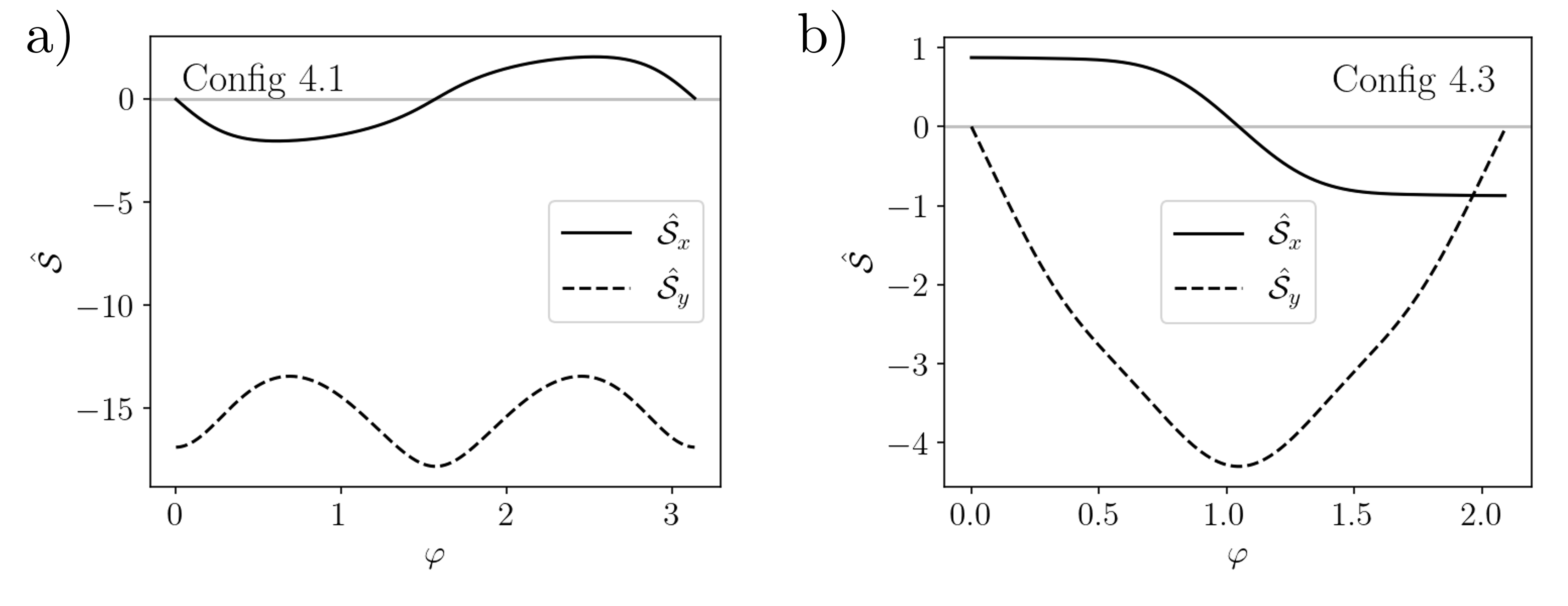}
    \caption{\textbf{Shafranov shift sensitivity to plasma $\beta$.} The plots show the shape gradient $\hat{\mathcal{S}}_{x}$ and $\hat{\mathcal{S}}_{y}$ for configurations 4.1 (a) and 4.3 (b) in the benchmark set. }
    \label{fig:shaf_shape_examples}
\end{figure}

We assess the Shafranov shift sensitivity in the configurations of the benchmark. Figure~\ref{fig:shaf_shape_examples} shows the shape gradient of the Shafranov shift for two different examples, and Table~\ref{tab:G_shaf_bench} summarises scalar features of the configurations.\footnote{The gradient calculation can be shown by finite differencing to be correct to machine precision. } 

\begin{table}
    \centering
    \begin{tabular}{c|c|c|c|c|c|c|c|}
       Configs &  4.1  & 4.2  & 4.3  & 4.4 (unopt)  & 4.4 (opt)  & \# 50  & \# 76  \\\hline
       $ \hat{\mathcal{S}}_\mathrm{max} $ & 17.82 & 2.52 & 4.30 & 7.02 & 6.02 & 2.40 & 3.31 \\
       $\beta_\Delta$ [\%] &  3.98 & 24.30 & 20.20 & 11.57 & 16.15 & 47.61 & 22.51 \\
       $\beta_c$ [\%] &  8.30 & 15.29 & 23.66 & 14.13 & 19.31 & 48.65 & 22.47 \\
       $|\iota_0|$ & 0.110 & 1.230 & 0.768 & 0.385 & 0.359 & 0.943 & 0.918 \\
       
    \end{tabular}
    \caption{\textbf{Shafranov shift sensitivity to plasma pressure.} The table shows the values of the  maximum sensitivity of the Shafranov shift, the estimate and umerical critical $\beta$ as well as the reference rotational transform on axis. The half-helicity fields appear to be the most resilient in the benchmark set. The comparison of $\beta_c$ to $\beta_\Delta$ show that although $\beta_c$ includes some key elements of the full phenomena, it can deviate significantly.}
    \label{tab:G_shaf_bench} 
\end{table}

The sensitivity $\mathcal{S}_\mathrm{max}$ shows that there exists a wide range of sensitivities to changes in plasma $\beta$. The gradient as a function of $\varphi$ delves deeper into these differences showing the stark difference in the binormal shift of the surfaces (which the half helicity field 4.3 appears to avoid to a large extent). The gradients also vanish at certain points in $\varphi$, corresponding to the directions that would break the up-down symmetry of the cross-sections at stellarator symmetric points. From this benchmark, the lowest number of field period configuration (config. 4.1) is the most sensitive, which also has the lowest value of rotational transform. This is true of $\beta_\Delta$ as well, showing that half-helicity configurations (namely configs. 4.3, \#50 and \#76) consistently exhibit a larger beta limits. 

The tools introduced here make it possible, in principle, to explore the generality of such observations, but an in depth analysis is left for future work. 

\section{Conclusions}

Recent works have demonstrated that it is possible to find approximately quasi-isodynamic stellarator equilibria to second order using near axis theory, and faithfully reproduce their properties in global equilibria at finite aspect ratio.  Such ``directly constructed'' stellarator equilibria can in principle be used as the basis for further integrated optimization to develop new stellarator designs.

However, to make the near-axis construction really practically useful, both for developing fundamental understanding, and as a tool for stellarator optimization, it is critical to develop techniques to navigate the space of second order solutions.  The present work tackles this task by defining a set of measures designed to assess the key properties that a quasi-isodynamic stellarator strives for, including low neoclassical transport, quality of omnigeneity, and robust stability.  Other measures already implemented (although not mentioned in the text) include the coil complexity proxy $L_{\nabla B}$ \citep{landreman2021a,Kappel_2024}, the effective Rosenbluth-Hinton residual \citep{rosenbluth1998poloidal,rodriguez2024zonal,zhu2025collisionless} and proxies to evaluate the maximum-$J$ property \citep{rodriguez2024maximum}.  Virtually any other measure that depends on magnetic geometry, for instance MHD ballooning stability or micro-turbulence, can be readily implemented as well in the future.

The procedure for obtaining fields to second order involves fixing a set of free functions at zeroth, first, and second order.  Unsurprisingly, there is a certain interaction between these choices, and the metrics have been formulated in a manner to help assess these choices.  We also propose constructive approaches to make these choices, {\em e.g.} a way to obtain a minimally shaped, marginally stable (in the magnetic well sense) field or to construct an optimally ``omnigenised'' configuration given a 1st order construction.

The toolset presented here, consisting of both absolute measures ({\em e.g.} $\epsilon_\mathrm{eff}$ on edge) and sensitivity measures (shape gradient of magnetic well) now opens the way to exploratory studies of the broad parameter space of QI stellarators, and systematic investigation of ``trade-offs'' -- {\em i.e.}, the compatibility of desired stellarators properties, underlying reasons, and strategies for effectively striking compromises.


\section*{Data availability}
The data and scripts that support the findings of this study are openly available at the Zenodo repository with DOI/URL 10.5281/zenodo.15340967.

\section*{Declaration of interests}
The authors report no conflict of interest.

\section*{Acknowledgements}
We thank Carolin N\"{u}hrenberg for her assistance with magnetic well definitions.

 \section*{Funding}
E. R. was partially supported by a grant by Alexander-von-Humboldt-Stiftung, Bonn, Germany, through a postdoctoral research fellowship. 

\appendix

\section{Benchmark configuration summary} \label{app:bench}
Throughout the paper we use a number of near-axis fields as a way of benchmark or illustration of the various measures introduced. In this section we summarise what those configurations consist of, indicating their main features and referring to the right piece of literature where appropriate.

\subsection{QI database} \label{app:QI_database}
In Section~\ref{sec:dB} we use a large set of QI configurations as a way to benchmark the $\delta\!B$ measure. This benchmark set had previously been used in work like \cite{rodriguez2024zonal}, and we now present some of its details. A full description of its construction and the theoretical elements that go into it will be the focus of future publication \citep{plunk2025-geometric}. For a specific example, see \cite{plunk2024back}.
\par
The benchmark set consists of first order QI configurations of number of field periods ranging $N=1$ to $6$, constructed as a three-parameter family as follows. The first parameter is related to the shape of the magnetic axis, which is described by its curvature, $\kappa$, and torsion, $\tau$.  For technical reasons, these must be defined differently for the case $N=1$, as compared to $N \geq 2$.  For the former we define
\begin{align}
    \kappa =  \cos^2\left(\frac{N \ell}{2}\right) \sin\left(\frac{N \ell}{2}\right)\left( \kappa_1 \sin(N \ell) + \kappa_2 \sin(2 N \ell) \right),\\ \tau = \tau_0 + \tau_1 \cos( N \ell) + \tau_2 \cos(2 N \ell),
\end{align}
while in the latter case ($N \geq 2$) we set $\kappa_2 = 0$ and $\tau_2 = 0$.
This difference is because four degrees of freedom are here used to close the curve smoothly for the case $N =1$, while only two degrees of freedom are required for $N \geq 2$.  Thus, a single parameter remains to define a one-parameter family of curves. The variable $\ell$ is the length along the curve. This describes a stellarator symmetric axis with zeroes of second order at the tops and third order at the bottom. This is consistent with a curve of half-helicity \citep{rodriguez2024near,Camacho2023helicity}. The total length of the axis is taken to be $L=2\pi$. 
\par
The magnetic field on axis $B_0$ is chosen to match the zeroes of $\kappa$,
\begin{equation}
  B_0 = 1 + 0.25 \cos(N \ell) + 0.0625 \cos(2N \ell),\label{eq:B0}
\end{equation}
and to have a mirror ratio of $\sim24\%$.\footnote{Here mirror ratio is defined as $\Delta_\mathrm{mirr}=(B_{0,\mathrm{max}}-B_{0,\mathrm{min}})/(B_{0,\mathrm{max}}+B_{0,\mathrm{min}})$, which is the most used form.} The breaking of omnigeneity necessary at first order is done through a smooth buffer region as described in \cite[Sec.~B.2.3]{rodriguez2024near}, of order $k=3$.
\par
The first order construction is finalised by the choice of the \textit{elongation-profile},
\begin{equation}
    \rho = \rho_0 + \rho_1 \cos(N \varphi), \label{eq:elongation-input-form}
\end{equation}
which is directly related to the more common near-axis quantities of \cite{landreman2019} and \cite{rodriguez2024near} by $\rho = \bar{e} + (1 + \sigma^2)/\bar{e}$ where $\bar{e}=\bar{d}^2B_0/\bar{B}$ (which is intimately related to the true elongation of the flux surface) \citep{plunk2024back}. These two parameters $\{\rho_0,\rho_1\}$, together with $\tau_1$ are our three parameters that span the QI database set considered.
\par
Considering different values of these parameters, the QI database is constructed with a total of 1680 configurations distributed in $\{63,502,358,244,250,263\}$ between number of field periods $\{1,2,3,4,5,6\}$. Further details on the procedure followed to construct such configurations will follow in a future paper.

\subsection{Benchmark configurations from \cite{rodriguez2024near}}
\begin{figure}
    \centering
    \includegraphics[width=\linewidth]{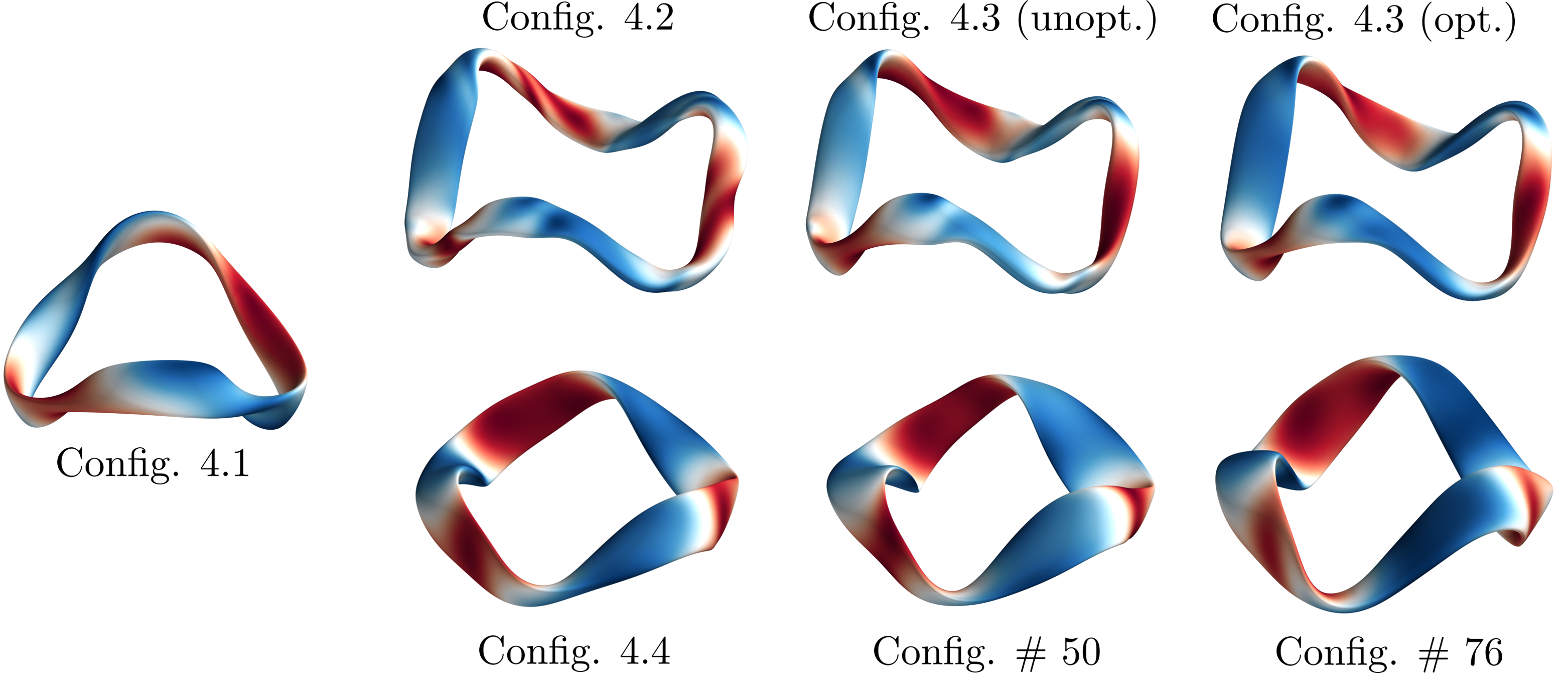}
    \caption{\textbf{3D rendition of configurations in the benchmark set.} 3D rendition of configurations in the benchmark set constructed for $r=0.1$ and with the colormap denoting the strength of the magnetic field $|\mathbf{B}|$.}
    \label{fig:bench_configs}
\end{figure}

In the rest of the paper we use a reduced set of benchmark configurations, mainly based on those used and explored in \cite[Sec.~4]{rodriguez2024near} to test the correct near-axis construction to 2nd order. Here we briefly summarise the key properties of each of those configurations,
\begin{itemize}
    \item \textbf{Configuration 4.1}: minimally shaped 2nd order stellarator symmetric, vacuum configuration with $N=2$, a smooth buffer region with $k=5$, mirror ratio $15\%$, and first order zeroes of curvature (zero helicity axis). 
    \item \textbf{Configuration 4.2}: shaped 2nd order stellarator symmetric, finite beta and current configuration with $N=3$, a smooth buffer region with $k=5$, mirror ratio $25\%$, and first order zeroes of curvature (zero helicity axis). 
    \item \textbf{Configuration 4.3}: minimally shaped 2nd order stellarator symmetric, vacuum configuration with $N=3$, a smooth buffer region with $k=5$, mirror ratio $25\%$, and third order zero of curvature at the bottom, and two at the top (half-helicity axis). 
    \item \textbf{Configuration 4.4}: minimally shaped 2nd order stellarator symmetric, vacuum configuration with $N=3$, a smooth buffer region with $k=5$, mirror ratio $25\%$, and third order zero of curvature at the bottom, and first order zeroes of curvature (zero helicity axis). The first order choices were then optimised to minimise the second order QI error in the central $20\%$ of the toroidal domain. In this paper we include both the unoptimised and optimised versions.
\end{itemize}
Figure~\ref{fig:bench_configs} shows 3D renditions of these configurations for reference.

\subsection{Additional half-helicity configurations}
In addition to the above, we also include two additional half-helicity fields constructed much in the same way as Config.~4.4, but corresponding to other constructions in the same parameter family,
\begin{itemize}
    \item Config. \#50: $\kappa_1=-7.451287$, $\tau_0=1.379097$, $\tau_1=-0.444444$, $\rho_0=4.5$, $\rho_1=-0.9$
    \item Config. \#76: $\kappa_1=-8.938611$, $\tau_0=1.284143$, $\tau_1=-0.177778$, $\rho_0=4.5$, $\rho_1=-0.9$
\end{itemize}
. The case of \# 76 is particularly interesting, as it has a rather omnigeneous intrinsic second order behaviour. See Figure~\ref{fig:bench_configs} for the configurations.

\section{Details on the near-axis form of $\epsilon_\mathrm{eff}$} \label{app:eps_eff}
In this Appendix we present the details in the asymptotic treatment of the effective ripple of a magnetic field with poloidal contours of $|\mathbf{B}|$ within the near-axis framework. The expression we must evaluate asymptotically is the following,
\begin{equation}
    \epsilon_\mathrm{eff}^{3/2}=\frac{\pi}{8\sqrt{2}}\frac{(\bar{R}\bar{B})^2}{\mathcal{G}^2}\,\int_{1/B_\mathrm{max}}^{1/B_\mathrm{min}}\lambda \hat{\mathcal{E}}(\lambda)\,\mathrm{d}\lambda, \tag{\ref{eqn:def_eps_eff_nae}}
\end{equation}
where,
\begin{subequations}
\begin{align}
    \hat{\mathcal{E}}(\lambda) &= \frac{1}{\pi}\int_0^{2\pi}\frac{\hat{H}(\lambda,\alpha)^2}{\hat{I}(\lambda,\alpha)}\,\mathrm{d}\alpha, \tag{\ref{eqn:E_nae}}\\
    \hat{H}(\lambda,\alpha) &= \frac{1}{\bar{B}}\int_{\varphi_-}^{\varphi_+}\frac{\mathcal{H}(\lambda,B)}{B^2}\mathbf{B}\times\nabla B\cdot\nabla\psi\,\mathrm{d}\varphi, \label{eqn:H_hat_nae} \\
    \hat{I}(\lambda,\alpha) &= \int_{\varphi_-}^{\varphi_+}\frac{\sqrt{1-\lambda B}}{(B/\bar{B})^2}\mathrm{d}\varphi, \label{eqn:I_hat_nae} \\
    \mathcal{G}^2 &= 2\left(\frac{1}{2\pi}\int_0^{2\pi}\mathrm{d}\alpha\int_0^{2\pi/N}\mathrm{d}\varphi\frac{|\nabla\psi|}{B^2}\right)^2\Bigg/\left(\frac{1}{2\pi}\int_0^{2\pi}\mathrm{d}\alpha\int_0^{2\pi/N}\frac{\mathrm{d}\varphi}{B^2}\right), \label{eqn:G_sq_nae} 
\end{align}  \label{eqn:eps_eff_fns_def}
\end{subequations}
\begin{equation}
    \mathcal{H}(\lambda,B)=\frac{\xiv}{(B/\bar{B})^2}\left(\frac{4}{\lambda B}-1\right). \tag{\ref{eqn:H_tilde}}
\end{equation}
where we have explicitly applied the considerations in Section~\ref{sec:eps_eff_adj} to write the expressions in the most near-axis friendly form.

\subsection{Expansion of $\hat{H}$}
For the expansion of $\hat{H}$, let us write using Boozer coordinates
\begin{equation}
    \mathbf{B}\times\nabla B\cdot\nabla\psi = \frac{B^2}{G+\iota I}(I\partial_\varphi-G\partial_\theta)B,
\end{equation}
which may be directly expanded,
\begin{align}
    \frac{\mathbf{B}\times\nabla B\cdot\nabla\psi}{B^2}\approx\, & rB_0d\cos(\alpha-\alpha_\mathrm{buf}) + \\
    & +r^2\left(B_{2s}^\mathrm{QI}\cos2\alpha-B_{2c}^\mathrm{QI}\sin2\alpha+\frac{I_2B_0'}{G_0}\right) + O(r^3),
\end{align}
where the order $O(r^3)$ term will include third order elements of the magnetic field magnitude, and thus we shall not include.  
\par
With this expansion of the radial drift, and noting that $\mathcal{H}$ vanishes at bounce points, the perturbation of $\hat{H}$ is a combination of the asymptotic form of the drift and that of the integrand $\mathcal{H}$. At first order, we easily obtain $\hat{H}\approx rh^{(1)}\sin\alpha$,
\begin{equation}
    h^{(1)} =\int_{\varphi_-}^{\varphi_+}\frac{B_0}{\bar{B}}\mathcal{H}(\lambda,B_0)d\sin\alpha_\mathrm{buf}\,\mathrm{d}\varphi, \tag{\ref{eqn:h_1}}
\end{equation}
where the odd part of the radial drift does not contribute, as its contributions from each side of the well precisely cancel out. 
\par
At second order we must consider both the contributions from the modification in the drift, but also the measure $\mathcal{H}$. To avoid the explicit appearance of the singular integrand $1/\sqrt{1-\lambda B_0}$, we integrate the resulting expression by parts (the boundary term being guaranteed to vanish for a field without puddles \citep{rodriguez2023higher}), and thus, after enforcing parity in $\varphi$ and collecting terms, $\hat{H}$ from both the perturbed $\hat{H}\approx rh^{(1)}\sin\alpha+r^2h^{(2)}\sin2\alpha$,
\begin{subequations}
\begin{align}
    h^{(2)} &= -2\int_{\varphi_-}^{\varphi_+}\mathcal{H}(\lambda,B_0)\frac{\Delta B_{2c}^\mathrm{QI}}{\bar{B}}\,\mathrm{d}\varphi,  \tag{\ref{eqn:h2}} \\
    \Delta B_{2c}^\mathrm{QI} &= B_{2c}^\mathrm{QI}-\frac{1}{4}\partial_\varphi\left(\frac{B_0^2d^2}{B_0'}\cos2\alpha_\mathrm{buf}\right). \tag{\ref{eqn:qi_2nd_order}}
\end{align}
\end{subequations}

\subsection{Expansion of $\hat{I}$}
We may now proceed in a similar fashion with the asymptotic evaluation of $\hat{I}$. It is useful to note that $\hat{I}$ is almost the expression for the second adiabatic invariant $\mathcal{J}_\parallel$ for trapped particles labelled by $\lambda$. Fortunately, the asymptotic treatment of such expressions (especially at higher order), with the careful handling of the boundary contributions, has already been done elsewhere, and thus we simply need to get the appropriate expressions from \cite{rodriguez2024maximum}.
\par
The leading order asymptotic form is simple, $\hat{I}\approx I^{(0)}$,
\begin{equation}
    I^{(0)} = \int_{\varphi_-}^{\varphi_+}\frac{\sqrt{1-\lambda B_0}}{(B_0/\bar{B})^2}\mathrm{d}\varphi. \tag{\ref{eqn:I0}}
\end{equation}
For the next orders, we can directly draw from Eq.~(A14) in \cite{rodriguez2024maximum}, and thus write separating $\hat{I}\approx I^{(0)}+rI^{(1)}\cos\alpha+r^2(\bar{I}^{(2)}+\tilde{I}^{(2)}\cos2\alpha)$,
\begin{equation}
    I^{(1)}=-2\int_{\varphi_-}^{\varphi_+}F^\star(\lambda,B_0)d\sin\alpha_\mathrm{buf}\,\mathrm{d}\varphi,
\end{equation}
and
\begin{subequations}
\begin{align}
    \bar{I}^{(2)}= & -2\int_{\varphi_-}^{\varphi_+}\frac{F^\star(\lambda,B_0)}{B_0}\left[B_{20}-\frac{1}{4}\partial_\varphi\left(\frac{B_0^2d^2}{B_0'}\right)\right]\,\mathrm{d}\varphi, \\
    \tilde{I}^{(2)}= & 2\int_{\varphi_-}^{\varphi_+}\frac{F^\star(\lambda,B_0)}{B_0}\left[B_{2c}^\mathrm{QI}-\frac{1}{4}\partial_\varphi\left(\frac{B_0^2d^2}{B_0'}\cos2\alpha_\mathrm{buf}\right)\right]\,\mathrm{d}\varphi,
\end{align}
\end{subequations}
where
\begin{equation}
    F^\star(\lambda,B_0)=\frac{1}{\sqrt{1-\lambda B_0}}\frac{1-3\lambda B_0/4}{(B_0/\bar{B})^2}.
\end{equation}
Note that all the $\alpha$ dependence of $\hat{I}$ vanishes in the QI limit, as it must given the similarity of $\hat{I}$ with $\mathcal{J}_\parallel$; the second adiabatic invariant is a flux function in an omnigeneous field \citep{bernardin1986}. The first order correction is purely driven by the buffer region, while the second comes from the second order QI mismatch. In that omnigeneous limit, $\bar{I}^{(2)}$ is the only higher order correction left to second order, and it represents the change in the field line length ($B_{20}$) but also in the velocity of the trapped particle along the bounce trajectory.

\subsection{Constructing $\mathcal{E}$}
Let us then consider now using the asymptotic definitions $\hat{H}\approx rh^{(1)}\sin\alpha+r^2h^{(2)}\sin2\alpha$ and $\hat{I}\approx I^{(0)}+rI^{(1)}\cos\alpha+r^2(\bar{I}^{(2)}+\tilde{I}^{(2)}\cos2\alpha)$,
\begin{align}
    \frac{1}{2\pi}\int_0^{2\pi}\frac{\hat{H}^2}{\hat{I}}\mathrm{d}\alpha\approx & \frac{r^2}{2}\left[\frac{(h^{(1)})^2}{I^{(0)}}+\right. \\
    & r^2\Bigg(\frac{(h^{(2)})^2}{I^{(0)}}\underbrace{-\frac{h^{(2)}h^{(1)}}{I^{(0)}}\frac{I^{(1)}}{I^{(0)}}}_{\textcircled{1}}+\underbrace{\frac{(h^{(1)})^2}{I^{(0)}}\Bigg[\Bigg(\frac{I^{(1)}}{2I^{(0)}}\Bigg)^2-\frac{\bar{I}^{(2)}-\tilde{I}^{(2)}/2}{I^{(0)}}\Bigg]}_{\textcircled{2}}\Bigg)\Bigg]. \label{eqn:B10}
\end{align}
A careful consideration of this expansion will show that we have a priori unfairly omitted a contribution that is also of order $r^4$, and that comes from beating of $h^{(1)}$ with what would be $h^{(3)}$, and in particular its $\sin\alpha$ harmonic. That term involves higher order magnetic field components, notably $B_{31}^c$, and thus goes beyond the second order considerations here. Setting this third component aside, the term will also have components inherited from lower orders, which can be evaluated explicitly to assess their magnitude. In practice, these contributions appear to be small. All in all, we shall ignore such contribution.
\par
In fact, in practice, at order $O(r^4)$ the main contribution comes from $h^{(2)}$; namely, the breaking of omnigeneity at second order. Thus, in practice, and as explicitly presented in the text, we will focus on this term.

\subsection{Expansion of geometric factor $\mathcal{G}$}
Let us write the geometric factor $\mathcal{G}^2=2D^2/L$ and consider the asymptotic expansion of the integrals. Following \cite[Eq.~(33)]{jorge2020naeturb} and carrying it out to higher order, we write
\begin{align}
    D=\frac{1}{2\pi}\int\mathrm{d}\alpha\int\frac{\mathrm{d}\varphi}{B^2}|\nabla\psi|\approx &\, r\overbrace{\frac{1}{2\pi}\int\mathrm{d}\alpha\int\frac{\mathrm{d}\varphi}{B_0}T_1}^{D_1}+r^2\overbrace{\frac{1}{2\pi}\int\mathrm{d}\alpha\int\frac{\mathrm{d}\varphi}{B_0}T_1\left(\frac{T_2}{T_1}-2\frac{B_1}{B_0}\right)}^{D_2}+\nonumber\\
    & +r^3\underbrace{\frac{1}{2\pi}\int\mathrm{d}\alpha\int\frac{\mathrm{d}\varphi}{B_0^2}T_1\left(\frac{T_3}{T_1}-2\frac{B_1T_2}{B_0T_1}+3\left(\frac{B_1}{B_0}\right)^2-2\frac{B_2}{B_0}\right)}_{D_3},
\end{align}
where $|\nabla\psi|/B_0\approx rT_1+r^2T_2+\dots$, and we may explicitly write
\begin{subequations}
    \begin{align}
        T_1 = & \sqrt{(\partial_\chi X_1)^2+(\partial_\chi Y_1)^2}, \\
        T_2 = & \frac{1}{T_1}\left[T_1^2\frac{B_1}{B_0}+\partial_\chi X_1\partial_\chi X_2+\partial_\chi Y_1\partial_\chi Y_2\right],
    \end{align}
\end{subequations}
and we omit the explicit form of $T_3$ for brevity. Note that $D_2\approx0$ as we have odd $\alpha$-harmonics that average to zero (even if there is a complicated $\alpha$ dependence in $T_1$). To evaluate the $\alpha$-average of the other quantities, we must explicitly write their poloidal angle dependence. For instance, $T_1$ )in the text Eq.~(\ref{eqn:G2_Psi1})), \cite[Eq.~(33)]{jorge2020naeturb}, 
\begin{equation}
    |\nabla\psi|^2\approx r^2B_0^2\left[\left(X_{1c}\sin\chi-X_{1s}\cos\chi\right)^2+\left(Y_{1c}\sin\chi-Y_{1s}\cos\chi\right)^2\right].
\end{equation}
Because we have $|\nabla\psi|$ instead of $|\nabla\psi|^2$ the integral over $\chi$ must be performed numerically (or related to elliptic integrals \citep[Eq.~(3.670.1)]{gradshteyn2014table}).  
\par
The expansion of $L$ is simpler, and only the expansion in $B$ is necessary. Considering the average over $\alpha$ to eliminate the bare harmonics of $\chi$, we write,
\begin{equation}
    L=\frac{1}{2\pi}\int\mathrm{d}\alpha\int\frac{\mathrm{d}\varphi}{B^2}\approx \underbrace{\int\frac{\mathrm{d}\varphi}{B_0^2}}_{L_0} + r^2\underbrace{\frac{1}{2\pi}\int\mathrm{d}\alpha\int\frac{\mathrm{d}\varphi}{B_0^2}\left[3\left(\frac{B_1}{B_0}\right)^2-2\frac{B_2}{B_0}\right]}_{L_2}.
\end{equation}
With both of these,
\begin{equation}
    \frac{1}{\mathcal{G}^2}\approx\frac{L_0}{G_0D_1^2}\bigg[1+r^2\underbrace{\bigg(\frac{L_2}{L_0}-\frac{G_2}{G_0} - 2\frac{D_3}{D_1}\bigg)}_{\textcircled{3}}\bigg].
\end{equation}
In practice, it is the leading order piece that matters, and which we defined as $\mathcal{G}^{(0)}$ in the main text.

\subsection{Assessment of approximation}
With the approximations and expansions considered above, we may then construct the leading order forms of $\epsilon_\mathrm{eff}^{3/2}$ as presented in Section~\ref{sec:eps_eff}. We had argued that some of the terms involved in the asymptotic expression of the effective ripple are, in practice, subsidiary, even though not in the $O(r)$ sense. We now provide some numerical evidence in such regard using the benchmark configurations in the paper as test-bed. In Table~\ref{tab:eps_eff_approx} we summarise the relative magnitude of different contributions in an attempt to argue the correctness of the approximation to $\epsilon_\mathrm{eff}$. We show by $\textcircled{1}$, $\textcircled{2}$ and $\textcircled{3}$ the relative contribution by the terms denoted by these symbols in the asymptotic expansions above to $\epsilon_\mathrm{eff}^{3/2,(2)}$ . The smallness of these terms gives us an argument to focus on the contribution by the second order QI breaking and use it as apart of our effective ripple measure. The benchmark in Figure~\ref{fig:eps_eff_bench} also seconds this.

\begin{table}
    \centering
    \begin{tabular}{c|c|c|c|c|c|c|c|}
       Configs &  4.1  & 4.2  & 4.3  & 4.4 (unopt)  & 4.4 (opt)  & \# 50  & \# 76  \\\hline
       $\textcircled{1}$ & 9.99E-06 & 2.37E-04 & -2.99E-04 & 4.80E-05 & 2.35E-05 & -8.30E-05 & 2.31E-04 \\
       $\textcircled{2}$ & -1.72E-06 & -2.69E-05 & -5.76E-05 & -1.66E-05 & -1.83E-05 & -7.11E-06 & -9.91E-05 \\ 
       $\textcircled{3}$ & 6.26E-08 & -1.85E-06 & 3.25E-06 & -3.31E-07 & -1.04E-06 & 1.98E-07 & 6.68E-06
    \end{tabular}
    \caption{\textbf{Relative magnitude of additional asymptotic terms in $\epsilon_\mathrm{eff}^{3/2,(2) }$.} The table shows the relative contribution to $\epsilon_\mathrm{eff}^{3/2,(2)}$ of the asymptotic terms $\textcircled{1}$, $\textcircled{2}$ and $\textcircled{3}$. The smallness of these contributions throughout the benchmark configurations supports, along with the role of QI breaking, the expression for the effective ripple measure used.}
    \label{tab:eps_eff_approx}
\end{table}

\section{Details of the ripple well measure $A_w$ calculation} \label{app:rw}
In this appendix we develop the formalism necessary to efficiently evaluate the ripple well measure $A_w$ presented in the main text, Section~\ref{sec:rw}; namely,
\begin{equation}
    A_w=R/\min\left\{r ~|~ \exists ~\theta,\varphi : \partial_\varphi|_\alpha B(r,\theta,\varphi)=0, \partial_\varphi^2|_\alpha B(r,\theta,\varphi)=0\right\}. \tag{\ref{eqn:r_w}}
\end{equation}
The multi-valued form of $A_w$ for each fixed $\varphi$ was defined as $\hat{A}_w$, with $A_w=\mathrm{max}_\varphi\hat{A}_w$.

\subsection{First order construction: too simple}
Let us consider first, as a way of introduction, $A_w$ for the simplest case of a first order field: $B(\theta,\varphi)=B_0(\varphi) + r B_1(\theta,\varphi)$, where $B_1(\theta,\varphi)$ is given in Eq.~(\ref{eqn:B1}). The two conditions in Eq.~(\ref{eqn:r_w}) to solve simultaneously are,
\begin{subequations}
    \begin{align}
        B_0'+r(S'\cos\alpha-C'\sin\alpha) = 0, \label{eqn:well_1st_order_I}\\
        B_0''+r(S''\cos\alpha-C''\sin\alpha) = 0, \label{eqn:well_1st_order_II}
    \end{align}
\end{subequations}
where $S = B_0d\sin\alpha_\mathrm{buf}$, $C = B_0d\cos\alpha_\mathrm{buf}$ and primes denote derivatives in $\varphi$. Combining these equations and assuming $r\neq0$, one may solve for $\tan \alpha$ and substitute it back to retain a real, positive root for $r$,
\begin{equation}
    R/\hat{A}^{(1)}_w=\left|\frac{B_0'}{S'-C'\tan\alpha}\sqrt{1+\tan^2\alpha}\right|, \label{eqn:rw_1st}
\end{equation}
where $\alpha$ is given by,
\begin{equation}
    \tan\alpha = \frac{(S'/B_0')'}{(C'/B_0')'},
\end{equation}
defined in the appropriate quadrant so that $1/\hat{A}_w^{(1)}$ solves the original set of equations. Note that in obtaining an expression for $\tan \alpha$ we have assumed that the equation is not trivially satisfied by $r=0$. This would have worked for $B_0'=0=B_0''$, which now instead sees $\hat{r}_w\rightarrow\infty$.\footnote{To show this it is sufficient to consider the no-buffer limit of Eq.~(\ref{eqn:rw_1st}), which at the bottom of the well requires $B_0''d'=0$ and $r\sin\alpha=B_0'/C'$, so that $r_w=B_0'/C'=\frac{1}{d}+\frac{B_0'}{B_0d'}$. Regardless of how one chooses the orders of $B_0$ and $d$, this expression diverges as the bottom of the well is approached.} This should make it clear that the appearence of secondary wells is not the same as the break-down of $|\mathbf{B}|$ contour topology.
\par
In practice, the appearence of secondary wells due to the first order variation of the field is too simplistic. It is far from capturing the evolution of a second order near-axis field construction. We must therefore turn to the more complex higher order consideration.

\subsection{Higher order form}
The procedure at higher orders resembles the procedure followed in the construction of the critical radius $r_c$ in \cite{landreman2021a}. The latter corresponds to the smallest distance from the axis at which flux surfaces become ill-behaved; namely, the first instance of a vanishing coordinate Jacobian $\mathcal{J}=\partial_\psi\mathbf{x}\times\partial_\theta\mathbf{x}\cdot\partial_\varphi\mathbf{x}=0$. That problem requires the solution of a set of equations formally analogous to those for $A_w$, see \cite[Sec.~4]{landreman2021a}. We exploit that analogy and adapt that work to our problem. 
\par
Using the helical angle $\chi=\theta-N\varphi$ where $N$ is the helicity of the magnetic axis \citep{rodriguez2022phases,Camacho2023helicity,rodriguez2024near}, and writing $B_1(\chi,\varphi)=B_{1c}(\varphi)\cos\chi+B_{1s}(\varphi)\sin\chi$ and $B_2(\chi,\varphi)=B_{20}(\varphi)+B_{2c}(\varphi)\cos2\chi+B_{2s}(\varphi)\sin2\chi$, the first condition that $\hat{A}_w$ must satisfy is $\partial_\varphi B|_\alpha=0$, which explicitly reads,
\begin{equation}
    g_0 + r\left(g_{1c}\cos\chi+g_{1s}\sin\chi\right)+r^2\left(g_{20}+g_{2c}\cos2\chi+g_{2s}\sin2\chi\right) = 0, \label{eqn:rw_eqI}
\end{equation}
where 
\begin{subequations}
    \begin{align}
        g_0 &= B_0', \\
        g_{1c} &= B_{1c}'+\bar{\iota}B_{1s}, &
        g_{1s} &= B_{1s}'-\bar{\iota}B_{1c}, \\
        g_{20} &= B_{20}', &
        g_{2c} &= B_{2c}'+2\bar{\iota}B_{2s}, &
        g_{2s} &= B_{2s}'-2\bar{\iota}B_{2c}.
    \end{align}
\end{subequations}
The second condition, namely $\partial_\varphi^2B|_\alpha=0$ can also be found explicitly and written in a form analogous to Eq.~(\ref{eqn:rw_eqI}). We shall not give it explicitly for brevity. Given these two equations, we approach their solution by isolating $r$ for every $\varphi$ and eliminating it from the equations, reducing the system to a single equation on harmonics of $\chi$. We shall call this the \textit{K-equation}. Following this procedure, we may write
\begin{equation}
    r=\frac{f_{1c}\cos\chi+f_{1s}\sin\chi}{f_{20}+f_{2c}\cos2\chi+f_{2s}\sin2\chi}, \label{eqn:r_eqn}
\end{equation}
where 
\begin{subequations}
    \begin{align}
        f_{1c} &= B_0'(B_{1c}''+2\bar{\iota}_0B_{1s}'-\bar{\iota}_0^2B_{1c}) - B_0''(B_{1c}'+\bar{\iota}_0B_{1s}), \\
        f_{1s} &= B_0'(B_{1s}''-2\bar{\iota}_0B_{1c}'-\bar{\iota}_0^2B_{1s}) + B_0''(-B_{1s}'+\bar{\iota}_0B_{1c}), \\
        f_{20} &= B_0''B_{20}'-B_0'B_{20}'', \\
        f_{2c} &= -B_0'(B_{2c}''+4\bar{\iota}_0B_{2s}'-4\bar{\iota}_0^2B_{2c}) + B_0''(B_{2c}'+2\bar{\iota}_0B_{2s}), \\
        f_{2s} &= B_0'(-B_{2s}''+4\bar{\iota}_0B_{2c}'+4\bar{\iota}_0^2B_{2s}) + B_0''(B_{2s}'-2\bar{\iota}_0B_{2c}).        
    \end{align}
\end{subequations}
Eliminating $r$ from Eq.~(\ref{eqn:rw_eqI}), we are left with the following \textit{K-equation} in the coordinates $\chi$ and $\varphi$,
\begin{equation}
    K_0+K_{2c}\cos2\chi+K_{2s}\sin 2\chi+K_{4s}\sin 4\chi+K_{4s}\sin 4\chi = 0,
\end{equation}
where
\begin{subequations}
    \begin{multline}
        K_0 = \frac{1}{4}\left[2g_0(f_{2c}^2 + 2 f_{20}^2 + f_{2s}^2) + f_{1c} (f_{2c} g_{1c} +  f_{2s} g_{1s} ) + f_{1s} (f_{2s} g_{1c} - f_{2c} g_{1s}) +\right. \\
        \left.2 f_{20} (f_{1c} g_{1c} + f_{1s} g_{1s}) + 2 (f_{1c}^2+f_{1s}^2) g_{20} + (f_{1c}^2 - f_{1s}^2) g_{2c} + 2 f_{1c} f_{1s} g_{2s}\right],
    \end{multline}
    \begin{multline}
        K_{2c} = \frac{1}{2}\left[f_{1c} f_{2c} g_{1c} + f_{20} (4 f_{2c} g_0 + f_{1c} g_{1c} - f_{1s} g_{1s}) + f_{1c}^2 (g_{20} + g_{2c}) + \right.\\
        \left.f_{1s} (f_{2c} g_{1s} - f_{1s} g_{20} + f_{1s} g_{2c})\right]
    \end{multline}
    \begin{multline}
        K_{2s} = \frac{1}{2}\left[f_{1c} f_{2s} g_{1c} + f_{1s} f_{2s} g_{1s} + f_{20} (4 f_{2s} g_0 + f_{1s} g_{1c} + f_{1c} g_{1s}) + \right.\\
        \left.2 f_{1c} f_{1s} g_{20} + (f_{1c}^2 + f_{1s}^2) g_{2s}\right]
    \end{multline}
    \begin{multline}
        K_{4c} = \frac{1}{4}\left[2 (f_{2c}^2 - f_{2s}^2) g_0 - f_{2s} (f_{1s} g_{1c} + f_{1c} g_{1s}) + f_{2c} (f_{1c} g_{1c} - f_{1s} g_{1s}) + \right. \\
        \left. (f_{1c}^2 - f_{1s}^2) g_{2c} - 2 f_{1c} f_{1s} g_{2s}\right],
    \end{multline}
    \begin{multline}
        K_{4s} = \frac{1}{4}\left[f_{2c} (4 f_{2s} g_0 + f_{1s} g_{1c} + f_{1c} g_{1s}) + f_{1c} (f_{2s} g_{1c} + 2 f_{1s} g_{2c}) + \right. \\
        \left.(f_{1c}^2-f_{1s}^2) g_{2s} - f_{1s} f_{2s} g_{1s}\right].
    \end{multline}
\end{subequations}
With the equation cast this way, we may then proceed the way that is detailed in \cite[Sec.~4.2]{landreman2021a}. That is, we solve the $K$-equation for $\sin2\chi$ which is a quartic, being careful about the sign of the trigonometric functions. As a result, we have up to four real roots for each value of $\varphi$ considered in the toroidal domain. Each one of these roots corresponds to a value of $r$ using Eq.~(\ref{eqn:r_eqn}), and thus represents a triplet $(r,\chi,\varphi)$. Finding $\alpha=\chi-\bar{\iota}\varphi$, we may then represent the values of $r$ for these multitude of roots as a function of $\alpha$, namely $\hat{A}_w$. This is what we represent in Figures~\ref{fig:rrw_examples} in Section~\ref{sec:rw}. The largest of all these roots is $A_w$.

\section{Linear system of equations at second order} \label{app:linear_system_2nd}
The construction of the near axis field at second order has been detailed elsewhere, most notably in \cite{landreman2019}, and in the QI scenario in \cite{rodriguez2024near}, where a thorough and pedagogical description is provided. Here we shall not reproduce the derivation of the construction, nor the way in which this is solved. However we shall for completeness write some of the key equations involved in the near-axis construction to 2nd order. In particular, the equations that the different pieces of the second order shaping satisfy, focusing on the operator notation that we use in this paper. 
\subsection{Expressions for $Y_2$}
The harmonics of $Y_2$ are defined through equations (A32)-(A33) of \citep{landreman2019} as,
\begin{equation}
    \begin{pmatrix} Y_{2c} \\ Y_{2s} \end{pmatrix} = \underbrace{\begin{pmatrix} \mathcal{Y}_{2c}^{0} \\ \mathcal{Y}_{2s}^{0} \end{pmatrix}}_{\pmb{\mathcal{Y}}_0} + \underbrace{\begin{pmatrix} \mathcal{Y}_{2c}^{X_{20}} & \mathcal{Y}_{2c}^{Y_{20}} \\ \mathcal{Y}_{2s}^{X_{20}} & \mathcal{Y}_{2s}^{Y_{20}} \end{pmatrix}}_{\bar{\mathsfbi{Y}}}\begin{pmatrix} X_{20} \\ Y_{20} \end{pmatrix} +
    \underbrace{\begin{pmatrix} \mathcal{Y}_{2c}^{X_{2c}} & \mathcal{Y}_{2c}^{X_{2s}} \\ \mathcal{Y}_{2s}^{X_{2c}} & \mathcal{Y}_{2s}^{X_{2s}} \end{pmatrix}}_{\hat{\mathsfbi{Y}}}\begin{pmatrix} X_{2c} \\ X_{2s} \end{pmatrix} \label{eqn:Y_eqs}
\end{equation}
where the components of each matrix are
\par
\begin{subequations}
\begin{minipage}{0.5\textwidth}
    \begin{align}
        \mathcal{Y}_{2c}^{0} &= \frac{\kappa\bar{B}}{B_0}\frac{X_{1c}X_{1s}}{X_{1c}^2+X_{1s}^2}, \\
        \mathcal{Y}_{2c}^{X_{20}} &= \frac{X_{1s}Y_{1s}-X_{1c}Y_{1c}}{X_{1c}^2+X_{1s}^2}, \\
        \mathcal{Y}_{2c}^{Y_{20}} &= -1+2\frac{X_{1c}^2}{X_{1c}^2+X_{1s}^2},  \\
        \mathcal{Y}_{2c}^{X_{2c}} &= \frac{X_{1c}Y_{1c}+X_{1s}Y_{1s}}{X_{1c}^2+X_{1s}^2} = \mathcal{Y}_{2s}^{X_{2s}}, 
    \end{align}
\end{minipage}
\hfill
\begin{minipage}{0.5\textwidth}
\begin{align}
        \mathcal{Y}_{2s}^{0} &= \frac{\kappa\bar{B}}{2B_0}\frac{X_{1s}^2-X_{1c}^2}{X_{1c}^2+X_{1s}^2}, \\
        \mathcal{Y}_{2s}^{X_{20}} &= -\frac{X_{1s}Y_{1c}+X_{1c}Y_{1s}}{X_{1c}^2+X_{1s}^2}, \\
        \mathcal{Y}_{2s}^{Y_{20}} &= 2\frac{X_{1c}X_{1s}}{X_{1c}^2+X_{1s}^2}, \\
        \mathcal{Y}_{2c}^{X_{2s}} &= \frac{X_{1c}Y_{1s}-X_{1s}Y_{1c}}{X_{1c}^2+X_{1s}^2}=-\mathcal{Y}_{2s}^{X_{2c}},
    \end{align}
\end{minipage}
\end{subequations}
\par
This is a purely algebraic equation, which involves no derivatives. 

\subsection{Equation for $X_{20}$ and $Y_{20}$} \label{app:LfF}
To complete the construction at second order, using $X_{2c}$ and $X_{2s}$ as inputs to the problem, we need to solve for $X_{20}$ and $Y_{20}$. The equations to solve are (A41)-(A42) of \cite{landreman2019}, which constitute two coupled first order ODEs in $\varphi$ to be solved simultaneously. Explicitly solving such a system requires rewriting all second order quantities explicitly in terms of $X_{20},Y_{20},X_{2c},X_{2s}$ and their derivatives, which involves Eq.~(\ref{eqn:Y_eqs}). 
\par
Instead of writing all the components of these operators out explicitly, we show how one should systematically proceed to find these through one particular example. The fully fleshed components may be found in the numerical implementation of the near-axis construction presented in the repository published alongside this work. In its operator form, the full expression is
\begin{equation}
    \hat{\mathbb{L}}\begin{pmatrix} X_{20} \\ Y_{20} \end{pmatrix} = -\hat{\mathbb{F}}\begin{pmatrix} X_{2c} \\ X_{2s} \end{pmatrix} + \pmb{f}, \label{eqn:LFf}
\end{equation}
and now we focus on constructing explicitly, say, $\mathbb{L}_{00}$. That is, the piece of the operator acting on $X_{20}$ in Eq.~(A41) of \cite{landreman2019}, not just explicitly, but also through other 2nd order functions. Here both $\hat{\mathbb{L}}$ and $\hat{\mathbb{F}}$ represent linear differential operators of first order that act on a vector of dimension 2, as shown. The vector $\pmb{f}$ represents combination of first order terms.
\par
Let us be more explicit, and consider the contributions one by one. First, the explicit $X_{20}'$ components,
\begin{subequations}
\begin{equation}
    A_{X_{20}'}=-X_{1s}\frac{\mathrm{d}}{\mathrm{d}\varphi}, 
\end{equation}
and explicit in $X_{20}$,
\begin{equation}
    A_{X_{20}}=-Y_{1s}\left(\tau\ell'-2\frac{I_2}{\bar{B}}\ell'\right)-4Y_{1c}\frac{G_0}{\bar{B}}Z_{2c}-4Y_{1s}\frac{G_0}{\bar{B}}Z_{2s}.
\end{equation}
We must then also consider the pieces that depend on $Y_{2c}$ and $Y_{2s}$, which we know are also related to $X_{20}$ by Eq.~(\ref{eqn:Y_eqs}). So we write again,
\begin{align}
    A_{Y_{2c}}=4X_{1s}\frac{G_0}{\bar{B}}Z_{2s}-X_{1c}\left(4\frac{G_0}{\bar{B}}Z_{20}-2\beta_0\ell'\right)+X_{1s}\left(\tau\ell'-2\ell'\frac{I_2}{\bar{B}}\right)-2Y_{1c}\bar{\iota}_0 \\
    A_{Y_{2s}}=-4X_{1s}\frac{G_0}{\bar{B}}Z_{2c}-X_{1c}\left(\tau\ell'-2\ell'\frac{I_2}{\bar{B}}\right)-X_{1s}\left(4\frac{G_0}{\bar{B}}Z_{20}-2\beta_0\ell'\right)-2Y_{1s}\bar{\iota}_0,
\end{align}
and finally the terms that depend on the derivatives of these,
\begin{align}
    A_{Y_{2c}'}=-Y_{1s}\frac{\mathrm{d}}{\mathrm{d}\varphi}, \\
    A_{Y_{2s}'}=Y_{1c}\frac{\mathrm{d}}{\mathrm{d}\varphi}.
\end{align}
\end{subequations}
With these expressions, we may then write $\hat{\mathbb{L}}_{00}$ explicitly,
\begin{equation}
    \hat{\mathbb{L}}_{00} = A_{X_{20}}+\mathcal{Y}_{2c}^{X_{20}}A_{Y_{2c}}+\mathcal{Y}_{2s}^{X_{20}}A_{Y_{2s}}-Y_{1s}(\mathcal{Y}_{2c}^{X_{20}})'+Y_{1c}(\mathcal{Y}_{2s}^{X_{20}})'-(X_{1s}+Y_{1s}\mathcal{Y}_{2c}^{X_{20}}-Y_{1c}\mathcal{Y}_{2s}^{X_{20}})\frac{\mathrm{d}}{\mathrm{d}\varphi}.
\end{equation}
Following a similar procedure with the other components (which we spare the reader from) the remainder terms of $\hat{\mathbb{L}}$ may be found, as well as those of $\hat{\mathbb{F}}$ and $\pmb{f}$.

\subsection{Tokamak limit simplification} \label{app:2nd_order_simp_tok}
Even though the topology of $|\mathbf{B}|$ contours over flux surfaces is different in tokamaks compared to QI configurations, the equilibrium construction remains largely the same. For the tokamak limit expressions evaluated in the main text it is then appropriate to consider the tokamak limit of the operators above. The toroidal coordinate $\varphi$ being one of symmetry simplifies the evaluation of the expressions significantly, especially because the toroidal derivatives may be taken to vanish exactly.
\par
In the near-axis tokamak notation, we also have $X_{1c}=\bar{d}$ and $X_{1s}=0$, with $Y_{1s}=1/\bar{d}$, and defining the major radius $R_0=G_0/B_0$. Taking this into consideration, and going through the equations, the following is true in the tokamak limit. The operator $\hat{\mathbb{L}}$ becomes,
\begin{equation}
    \hat{\mathbb{L}}_\mathrm{tok}=-\bar{\iota}_0\frac{B_0}{\bar{d}}
    \begin{pmatrix}
        \bar{d}^2-\frac{3}{\bar{d}^2}(1+\sigma^2) & 4\sigma \\
        0 & 2
    \end{pmatrix}, \label{eqn:L_tok}
\end{equation}
which becomes diagonal in the case of an up-down symmetric tokamak (i.e. $\sigma=0$). Note here the possibility of a singular matrix if, in the up-down symmetric scenario, $\bar{d}^4=3$. This singularity was discussed in the main text and is directly related to how the equilibrium problem is being solved. This is however seldom a problem given the typical values of elongation considered. 
\par
For $\hat{\mathbb{F}}$,
\begin{equation}
    \hat{\mathbb{F}}_\mathrm{tok} = -\bar{\iota}_0\frac{3B_0}{\bar{d}^3}\begin{pmatrix}
        \bar{d}^4-1+\sigma^2 & 2\sigma \\
        2\sigma & -(\bar{d}^4-1+\sigma^2)
    \end{pmatrix}, \label{eqn:F_tok}
\end{equation}
which is a symmetric matrix.  Finally,
\begin{equation}
    \pmb{f}_\mathrm{tok}=\frac{B_0\bar{\iota}_0}{\bar{d} R_0}\begin{pmatrix}
        \frac{1}{2}[1+2(\bar{d}^4+\sigma^2)]-\left(\frac{\bar{d}}{\bar{\iota}_0}\right)^2\frac{\mu_0 p_2}{B_0^2/2} \\
        \frac{5}{2}\sigma
    \end{pmatrix}.
\end{equation}
The matrix operators can be easily inverted and thus used for the shape gradient calculations in the paper. 

\subsection{Triangularity}
It is often illuminating to use more geometrical definitions of the shaping other than those directly involved in the near-axis construction. At second order, we may define an approximation to triangularity following \cite{rodriguez2023mhd},
\begin{equation}
    \delta=2\left(\frac{Y_{2s}}{Y_{1s}}-\frac{X_{2c}}{X_{1c}}\right),
\end{equation}
which through Eq.~(\ref{eqn:Y_eqs}) depends on all $X_{20},X_{2s},X_{2c}$ and $Y_{20}$. For the purpose of this paper we are interested on the variations of this expression respect to the second order shape choices. That is,
\begin{equation}
    \delta\!\delta = -\frac{2}{Y_{1s}}\left[\begin{pmatrix} \mathcal{Y}_{2s}^{X_{20}} \\ \mathcal{Y}_{2s}^{Y_{20}}   \end{pmatrix}^T\hat{\mathbb{L}}^{-1}\hat{\mathbb{F}} - \begin{pmatrix} \mathcal{Y}_{2s}^{X_{2c}} \\ \mathcal{Y}_{2s}^{X_{2s}}   \end{pmatrix}^T\right]\begin{pmatrix} \delta\!X_{2c} \\ \delta\!X_{2s} \end{pmatrix} - \frac{2}{X_{1c}}\delta\!X_{2c},
\end{equation}
which is rather complicated. In the tokamak limit, using the expressions in Eqs.~(\ref{eqn:L_tok}) and (\ref{eqn:F_tok}), the expression simplifies significantly to give,
\begin{equation}
    \delta\!\delta_\mathrm{tok}=\frac{2}{\bar{d}}\frac{\bar{d}^4+3}{\bar{d}^4-3}\delta\!X_{2c}, \label{eqn:dd_dX2c}
\end{equation}
as used in the main text. Note the appearence of the diverging denominator here, which signals the huge variation in the geometric triangularity due to a change in the input function $X_{2c}$ when close to the resonance, emphasising the perilous balance there.

\section{General sensitivity theory} \label{app:grad}
In this appendix we consider the evaluation of the shape gradient of some functional $S$ with respect to the shaping degrees of freedom at second order. Let such a functional be of the form,
\begin{equation}
    S[B_{20},B_{2c},B_{2s}]=\int_0^{2\pi} f(B_{20},B_{2c},B_{2s},\varphi)\,\mathrm{d}\varphi,
\end{equation}
functional that depends directly on the second order magnetic field functions. As we change the second order shaping, we are interested in how $S$ changes, call it $\delta\!S$. Assuming $f$ to be differentiable with respect to the three components of the second order magnetic field magnitude,
\begin{equation}
    \delta\!S=\int_0^{2\pi} \left(\frac{\partial f}{\partial B_{20}}\delta\!B_{20}+\frac{\partial f}{\partial B_{2c}}\delta\!B_{2c}+\frac{\partial f}{\partial B_{2s}}\delta\!B_{2s}\right)\,\mathrm{d}\varphi,
\end{equation}
and the partial derivatives are meant to be taken keeping the other second order magnetic field components fixed\footnote{It truly is a variation in the functional theory sense.}. In order to proceed further, we must relate the variation of these magnetic field components to the shaping. To that end, we draw from the presentation of the second order construction in Section~2.4 of RPJ \citep{rodriguez2024near}, and the explicit equations in Appendix~A of LS \citep{landreman2019}.
\par
The shaping freedom is on the functions $X_{2s}$ and $X_{2c}$, which represent in some form the triangularity that we may exploit to modify $S$. More precisely, they describe the $m=2$ variation of the distance of the flux surfaces from the magnetic axis along the normal of the latter. As the flux surface is deformed, the magnetic field magnitude must change accordingly, and thus $B_{2c}$ and $B_{2s}$ are quite directly modified. From Eq.~(2.7) in RPJ or Eqs.~(A35-36) in LS, 
\begin{align}
    \delta\! B_{2c}=\kappa B_0\, \delta\! X_{2c}, \\
    \delta\! B_{2s}=\kappa B_0\, \delta\! X_{2s}.
\end{align}
The larger the curvature of the magnetic axis, the stronger the supported perpendicular gradient of the magnetic field, and thus the easier the magnitude of the magnetic field can be shaped. 
\par
The variation of $B_{20}$ (Eq.~(A34) in LS) depends directly on $X_{20}$, which must be self-consistently solved alongside $Y_{20}$ in Eq.~(\ref{eqn:LFf}). The dependence on $X_{2c}$ and $X_{2s}$ (and their derivatives) do therefore involve the inversion of differential operators, which in the notation of Appendix~\ref{app:LfF}, 
\begin{equation}
    \delta\!B_{20}=-\kappa B_0\begin{pmatrix} 1 & 0 \end{pmatrix}\hat{\mathbb{L}}^{-1}\hat{\mathbb{F}}\begin{pmatrix} \delta\!X_{2c} \\ \delta\!X_{2s} \end{pmatrix}.
\end{equation}
Once again, we encounter the proportionality to $\kappa$, which denotes that partial inability of the shaping to modify the field especially near inflection points. However, this does not mean that the shaping at these points has no influence on $S$. Effects lose locality through the inversion of $\hat{\mathbb{L}}$. 
\par
With this we write, 
\begin{equation}
    \delta\!S=\int_0^{2\pi} \underbrace{\kappa B_0\left[-\begin{pmatrix} \frac{\partial f}{\partial B_{20}} & 0 \end{pmatrix} \hat{\mathbb{L}}^{-1}\hat{\mathbb{F}} + \begin{pmatrix} \frac{\partial f}{\partial B_{2c}} & \frac{\partial f}{\partial B_{2s}} \end{pmatrix}\right]}_{=\mathcal{G}^T}\delta\!X_2\,\mathrm{d}\varphi,
\end{equation}
where $\mathcal{G}$ can be though of as the shape gradient $\mathcal{G}=(\mathcal{G}_{2c}, \mathcal{G}_{2s})$ and $\delta\!X_2=(\delta\!X_{2c},\delta\!X_{2s})$. The function $\mathcal{G}$ can then be interpreted as the so-called \textit{shape-gradient} of $S$ respect to the shaping $X_2$. That is, it measures the amount by which an infinitesimal local variation of second order shaping would change the non-local value of $S$.
\par
A priori, and with all the quantities in $\mathcal{G}$ being known (they belong to the first order NAE), we could directly calculate it. However, it is in practice convenient to avoid having to invert $\hat{\mathbb{L}}$ explicitly\footnote{Avoiding explicit inversion of the matrix in solving a linear system tends to be numerically faster and more stable \citep[Chap.~28]{cormen2022introduction}.}. Instead, we may formulate the calculation of $\mathcal{G}$ through an adjoint problem. For simplicity, let us discretise the equation, as we will do anyways in practice. Defining a grid $\{\varphi_0,\varphi_1,\dots, \varphi_{N-1}\}$, and discretising integrals with the appropriate numerical scheme weight $\int f\mathrm{d}\varphi\approx\sum w_if(\varphi_i)$, we write,
\begin{equation}
    \delta\!S = \sum_{i=0}^{2(N-1)}w_i\mathcal{G}_i\delta\!X_{2,i},
\end{equation}
where 
\begin{equation}
    \mathcal{G}_i=-\frac{1}{w_i}\left(\mathbb{F}^T\right)_{ij}\pmb{y}_j+\kappa B_0 \begin{pmatrix} \frac{\partial f}{\partial B_{2c}} \\ \frac{\partial f}{\partial B_{2s}} \end{pmatrix},
\end{equation}
and $\pmb{y}$ is the solution to the following adjoint problem,
\begin{equation}
    \mathbb{L}^T\pmb{y} =  \begin{pmatrix} \kappa B_0 w \frac{\partial f}{\partial B_{20}} \\ 0 \end{pmatrix}.
\end{equation}
Thus, to compute the shape gradient $\mathcal{G}$, the key is to perform an adjoint solve to find $\pmb{y}$. The system to solve is nevertheless almost the same as for the direct second order solve problem, Eq.~(\ref{eqn:LFf}), as it involves $\mathbb{L}$, and thus it involves a complexity analogous to a second order solve itself. This construction presented in generality may be used for a variety of different measures.

The answer will depend on how we discretise the integral, and thus on the weights $w$. This is the price to pay for not having to integrate by parts. In its most basic form, using the standard trapezoidal scheme, the weight \citep[pg.~202]{suli2003introduction}
\begin{equation}
    w_i=\begin{cases}
        \frac{1}{2}\left(P+\varphi_1-\varphi_{N-1}\right), \quad (i=0) \\
        \frac{1}{2}\left(\varphi_{i+1}-\varphi_{i-1}\right), \quad (0<i<N)
    \end{cases} \label{eqn:weight_trapezoid}
\end{equation}
where $P$ is the period of the domain ($2\pi$ if the whole toroidal angle extent is considered). 
\par
\subsection{Tokamak limit} \label{app:grad_tok}
Let us consider the simplifying limit of a tokamak, where the toroidal derivatives vanish, and thus the procedure is rather simple (in fact, without the need of using a particular collocation grid). In this limit, using the expressions in Appendix~\ref{app:2nd_order_simp_tok}, we may write for the magnetic well,
\begin{equation}
    \mathcal{G}^W=-\kappa B_0 \frac{\partial f}{\partial B_{20}}(\mathbb{L}^{-1}\mathbb{F})^T\begin{pmatrix} 1 \\ 0 \end{pmatrix} = -3\kappa B_0 \frac{\partial f}{\partial B_{20}}\begin{pmatrix} \frac{\bar{d}^4-1-3\sigma^2}{\bar{d}^4-3(1+\sigma^2)} \\ -\frac{2\sigma(\sigma^2+\bar{d}^4)}{3(1+\sigma^2)-\bar{d}^4} \end{pmatrix} 
\end{equation}
Using the derivative of $f$ for $W$ explicitly, Eq.~(\ref{eqn:f_derivatives}), and taking the up-down symmetric limit for simplicity ($\sigma=0$), we are left with,
\begin{equation}
    \mathcal{G}^W = \frac{3r_\mathrm{ref}^2}{\pi R}\frac{\bar{d}^4-1}{\bar{d}^4-3} \begin{pmatrix} 1\\ 0 \end{pmatrix} .
\end{equation}
This is the expression presented in the main text, Eq.~(\ref{eqn:Vpp_tok}).

\section{Variational re-shaping problem} \label{app:variational_shape}
In this Appendix we show how the variational problem of Section~\ref{sec:sens_shaping} leads to a very particular form for the  shaping inputs at second order in the near-axis construction that allows us to reliably construct stable approximately QI fields. Let us rewrite here the equation in the main text,
\begin{multline}
    T[X_{2c},X_{2s}]=\frac{1}{2\pi}\int_0^{2\pi}\left(X_{20}^2+Y_{20}^2+\frac{X_{2c}^2+X_{2s}^2+Y_{2c}^2+Y_{2s}^2}{2}\right)\,\mathrm{d}\varphi-\\
    -\frac{\lambda}{2\pi}\left[\int_0^{2\pi}\left(\mathcal{G}_{2c}^{W}X_{2c}+\mathcal{G}^{W}_{2s}X_{2s}\right)\,\mathrm{d}\varphi+W_\mathrm{ref}\right], \tag{\ref{eqn:T_variational}}
\end{multline}
which we need to take variations of. We shall construct the variational considerations of the discretised form of the problem. We could do it of the continuous case, but that would involve integrating by parts, and ultimately, in practice, a discrete version is really needed.
\par
Let us then focus on writing $T$ in the discrete form, thereon a vector $\mathbf{v}=(a(\varphi),b(\varphi))$ is defined as an element of $\mathbb{R}^{2N}$, with $(a_0,a_1,\dots,a_{N-1},b_0,\dots,b_{N-1})$ where the subscripts correspond to the function evaluated at $\varphi_i$, the discretised $\varphi$-grid. Writing the weight of the discretised integral as $\mathsfbi{W}=\mathrm{diag}(\mathbf{w},\mathbf{w})$, where for the trapezoidal rule $\mathbf{w}$ is defined in Eq.~(\ref{eqn:weight_trapezoid}), and defining a convenient scalar product $\langle \mathbf{a},\mathbf{b}\rangle _\mathsfbi{W}=\mathbf{a}^T\mathsfbi{W}\mathbf{b}$, we may then write Eq.~(\ref{eqn:T_variational}) as,
\begin{multline}
    2\pi T[X_{2c},X_{2s}]= \frac{1}{2}\left\langle \vect{X_{2c}, X_{2s}},\vect{X_{2c},X_{2s}}\right\rangle _\mathsfbi{W} 
    + \left\langle \vect{X_{20}, Y_{20}},\vect{X_{20},Y_{20}}\right\rangle _\mathsfbi{W} 
    +\frac{1}{2}\left\langle \vect{Y_{2c}, Y_{2s}},\vect{Y_{2c},Y_{2s}}\right\rangle _\mathsfbi{W} \\
    -\lambda \left[\left\langle \vect{\mathcal{G}^{W}_{2c}, \mathcal{G}^{W}_{2s}},\vect{X_{2c},X_{2s}}\right\rangle _\mathsfbi{W} + W_\mathrm{ref}\right]. \label{eqn:}
\end{multline}
We must now take variations of these expressions with respect to the second order input functions $X_{2c}$ and $X_{2s}$. For the first term this is trivial, but for the next two we must take the variation through the equations that define first $X_{20}$ and $Y_{20}$, and then $Y_{2c}$ and $Y_{2s}$. For the former we need Eq.~(\ref{eqn:LFf}), and for the latter, both that and Eq.~(\ref{eqn:Y_eqs}). We may summarise these variations as,
\begin{equation}
    \vect{\delta\!X_{20},\delta\!Y_{20}}=-\underbrace{\hat{\mathbb{L}}^{-1}\mathbb{F}}_{\mathsfbi{X}}\vect{\delta\!X_{2c},\delta\!X_{2s}}, \quad \vect{\delta\!Y_{2c},\delta\!Y_{2s}}=\underbrace{\left(\hat{\mathsfbi{Y}}-\bar{\mathsfbi{Y}}\hat{\mathbb{L}}^{-1}\mathbb{F}\right)}_{\mathsfbi{Y}}\vect{\delta\!X_{2c},\delta\!X_{2s}},
\end{equation}
with all matrices defined in Appendix~\ref{app:linear_system_2nd}. 
\par
Expressing all the remaining terms explicitly in terms of the 2nd order input functions, the variation $\delta\!T$ can be written as,
\begin{equation}
    2\pi \delta\!T= \left\langle\mathsfbi{M}\vect{X_{2c},X_{2s}}-\pmb{\Lambda}-\lambda\,\pmb{G},\vect{\delta\!X_{2c},\delta\!X_{2s}}\right\rangle_\mathsfbi{W}, \label{eqn:T_variation_defs}
\end{equation}
where
\begin{subequations}
\begin{align}
    \pmb{G} =& \vect{\mathcal{G}^{W}_{2c}, \mathcal{G}^{W}_{2s}}, \\
    \mathsfbi{M}=&\mathsfbi{1}+2\mathsfbi{W}^{-1}\mathsfbi{X}^T\mathsfbi{W}\mathsfbi{X}+\mathsfbi{W}^{-1}\mathsfbi{Y}^T\mathsfbi{W}\mathsfbi{Y}, \\
    \pmb{\Lambda}=&\left[2\mathsfbi{W}^{-1}\mathsfbi{X}^T\mathsfbi{W}-\mathsfbi{W}^{-1}\mathsfbi{Y}^T\mathsfbi{W}\bar{\mathsfbi{Y}}\right]\mathbb{L}^{-1}\pmb{f}-\mathsfbi{W}^{-1}\mathsfbi{Y}^T\mathsfbi{W}\pmb{\mathcal{Y}}_0,
\end{align}
\end{subequations}
and it should be clear that $\mathsfbi{M}$ is a matrix of dimensions $2N\times2N$, while $\pmb{\Lambda}$ is a $2N$ vector. 
\par
We are then in a position to require the variation Eq.~(\ref{eqn:T_variation_defs}) to vanish for all variations, implying 
\begin{equation}
    \mathsfbi{M}\vect{X_{2c},X_{2s}}=\pmb{\Lambda}+\lambda\pmb{G}.
\end{equation}
Using this into the constraint equation leads to an expression for the Lagrange multiplier $\lambda$,
 \begin{equation}
     \lambda = -\frac{W_\mathrm{ref}+\left\langle\pmb{G},\mathsfbi{M}^{-1}\pmb{\Lambda}\right\rangle_\mathsfbi{W}}{\left\langle\pmb{G},\mathsfbi{M}^{-1}\pmb{G}\right\rangle_\mathsfbi{W}}. \label{eqn:lagrange_mult}
 \end{equation}
 With that, the required precise form of the second order shaping is,
 \begin{equation}
     \vect{X_{2c}^\mathrm{mhd},X_{2s}^\mathrm{mhd}} = \mathsfbi{M}^{-1}\left(\pmb{\Lambda}+\lambda\pmb{G}\right). \label{eqn:shape_mhd}
 \end{equation}
This is the answer we sought. 
\par
In order to gauge the computational complexity involved to reach the necessary shaping in Eq.~(\ref{eqn:shape_mhd}) we may consider roughly the operations that will be needed. Starting from the last equation, Eq.~(\ref{eqn:shape_mhd}), we here will need two linear system solves. These very operations may be recycled in the calculation of the Lagrange multiplier in Eq.~(\ref{eqn:lagrange_mult}). However we are hiding equally costly operations in the form of matrix multiplication and inversion (which are numerically of equal complexity \citep[Chap.~3]{golub2013matrix}\citep[Chap.~28]{cormen2022introduction}), which are needed to evaluate both $\mathsfbi{M}$ and $\pmb{\Lambda}$. To avoid taking explicit inverses especially to avoid numerical instability, the necessary operation could be broadcasted from right to left, so that operations on vectors are used at all times. However, in practice, we see that a na\"{i}ve implementation is well behaved. 

\section{Estimation of critical $\beta$} \label{app:shaf}
In Section~\ref{sec:shaf} we introduced a measure of a critical plasma $\beta$ that leads to flux surfaces in the near-axis field construction touching. In this Appendix we present the details of said calculation and the exact meaning of the elements that go into it. 

\subsection{Problem set-up}
The problem we want to describe is as follows. Consider a plane normal to the axis at some $\varphi$ value, as we shall describe the problem slice by slice. To first order in $r$, the shape of the flux surfaces in such plane correspond to ellipses. At second order, these ellipses are rigidly shifted as well as triangularly shaped. For the estimation in this Appendix we will simply ignore that triangular shaping, and model the cross-sections as ellipses that are rigidly shifted.
\par
Let us be a bit more precise. The ellipses are defined in the signed Frenet-Serret frame (where $X$ and $Y$ correspond to Cartesian coordinates aligned with the normal and binormal directions respectively) by,
\begin{subequations}
    \begin{align}
        X/r&=X_{1c}\cos\chi+X_{1s}\sin\chi, \\
        Y/r&=Y_{1c}\cos\chi+Y_{1s}\sin\chi.
    \end{align}
\end{subequations}
This defines an implicit ellipse, following \cite[Eq.~(B1)]{rodriguez2023mhd},
\begin{equation}
    (1+\sigma^2)X^2-2XY\sigma\pelon+\pelon^2Y^2=\frac{\bar{B}}{B_0}\pelon,
\end{equation}
where $\pelon=B_0\bar{d}^2/\bar{B}$. Note that this form is analogous to the quasisymmetric case in \cite{rodriguez2023mhd}, where $\pelon=(\eta/\kappa)^2$. This is true with the exception of the right-hand-side, due to the variation of the elliptical cross-section with the variation of the field strength $B_0$ in the current QI case. However, this means that the description of the ellipses is identical to that there. Namely, we may characterise an ellipse by its elongation $\mathcal{E}=\tan e$ defined as the ratio of major to minor radius, and $\theta$, the angle that the major axis makes with $X$. Explicitly,
\begin{equation}
    \sin 2e=\frac{2\pelon}{\pelon^2+1+\sigma^2}, \quad \tan 2\theta=\frac{2\sigma\pelon}{\pelon^2-(1+\sigma^2)}. \label{eqn:ellipse_features}
\end{equation}
We note that the inversion of the tangent must be carried out with some care to correctly capture the described physical angle. 
\begin{figure}
    \centering
    \includegraphics[width=\linewidth]{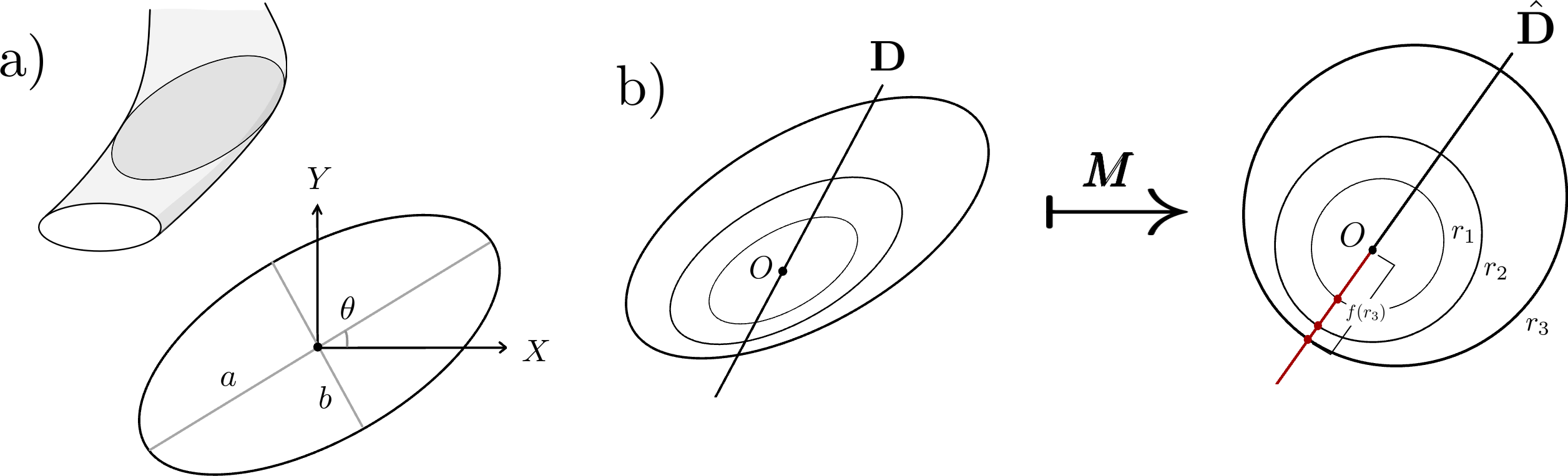}
    \caption{\textbf{Illustrating diagrams for the estimation of the critical $\beta$.} (a) Elliptical cross-section indicating the rotation angle $\theta$ and major and minoraxes, $a$ and $b$ respectively. (b) Shift of ellipses along the direction $\mathbf{D}$, and the $\mathbf{M}$-transformed scenario involving circles. The diagram shows the geometric meaning of $d$ as defined in Eq.~(\ref{eqn:inter_eqn}).}
    \label{fig:ellipses_shaf}
\end{figure}
\par
Now, we consider these ellipses to be shifted rigidly at second order, by an amount that we may refer to as,
\begin{equation}
    \mathbf{D}=r^2\begin{pmatrix} X_{20} \\ Y_{20}  \end{pmatrix},
\end{equation}
in the $(\hat{\pmb{X}}, \hat{\pmb{Y}})$ basis. Note that each value of $r$ defines a different ellipse. The goal of this appendix is to precisely describe when this displacement is such that two consecutive ellipses touch. 

\subsection{Making circles}
The problem of touching ellipses can be made significantly simpler by appropriately rotating and scaling the plane. First, we shall consider a clockwise rotation by $\theta$, the angle of the ellipse, Eq.~(\ref{eqn:ellipse_features}), so that all ellipses become `aligned' with the axes. By align we mean that the major and minor radii of all ellipses are parallel (or orthogonal) to the signed Frenet-Serret coordinate system. Then we scale the major axis direction by its magnitude and similarly for the minor radius, so that the result is a set of displaced circles of different radii. That is, we apply the linear map defined in this basis by,
\begin{equation}
    \pmb{M}=\begin{pmatrix}
        r/a & 0 \\
        0 & r/b
    \end{pmatrix} \begin{pmatrix}
        \cos\theta & \sin\theta \\
        -\sin\theta & \cos\theta
    \end{pmatrix}, \label{eqn:M_matrix}
\end{equation}
to all $(X,Y)$, so that ellipses are mapped to circles of radius $r$. Using $\mathcal{E}=a/b$ and the definition of the toroidal flux $\psi=\text{Area}\times B_0/2\pi$, so that $ab=\bar{B}r^2/B_0$, then
\begin{equation}
    a/r=\sqrt{\frac{\bar{B}}{B_0}\mathcal{E}}, \quad b/r=\sqrt{\frac{\bar{B}}{B_0\mathcal{E}}}. \quad
\end{equation}
This transformation may be applied to the shift $\mathbf{D}$, so that we have an effective distance $\hat{\mathbf{D}}=\pmb{M}\mathbf{D}$ (see Figure~\ref{fig:ellipses_shaf}).
\par
Through this transformation we have converted the problem of ellipses into a set of circles with varying radii displaced along the same direction by different amounts. The discussion of touching surfaces may thus be reduced to a discussion along this line.

\subsection{Touching circles}
Take the line on the rescaled plane defined by $\hat{\mathbf{D}}$ and define a function $f(r)$ that gives the distance from the origin to the point of intersection of the circle of radius $r$ with this line. Thanks to all being nicely stacked circles, 
\begin{equation}
    f(r) = r\left(1-r|\hat{\mathbf{D}}|\right), \label{eqn:inter_eqn}
\end{equation}
where we have defined the point of intersection in the halfline in which circles are bunching. 
\par
The point at which circles will touch corresponds to $\partial_rf=0$, which gives,
\begin{equation}
    A_\Delta=2R|\hat{\mathbf{D}}|, \label{eqn:A_Delta}
\end{equation}
where $A$ corresponds to the aspect ratio with $R$ the effective major radius defined in terms of the length of the magnetic axis. Below this aspect ratio, flux surfaces will intersect and the near-axis construction is invalid.

\subsection{Critical $\beta$}
The question is then how much does this critical aspect ratio change with the pressure in the problem. We know from the main text,
\begin{equation}
    \frac{\delta\!\mathbf{D}}{\delta\!\beta}=\frac{\hat{\pmb{\mathcal{S}}}}{r_\mathrm{ref}},
\end{equation}
where $\beta=\beta_2r_\mathrm{ref}^2$, and because it is linear in $\beta$, we may write exactly
\begin{equation}
    \mathbf{D}_\beta=\mathbf{D}_0+\frac{\delta\!\mathbf{D}}{\delta\!\beta}\delta\!\beta, \label{eqn:shaf_beta}
\end{equation}
where the subscript zero denotes the Shafranov shift at zero plasma beta.\footnote{If a known construction has a finite beta $\beta_0$, then one may simply modify the above with $\beta\mapsto\beta-\beta_0$. The problem is linear.} 
\par
Let us now define the critical $\beta_\Delta$ as the value of $\beta$ at which the touching of surfaces occurs at a reference aspect ratio $A_\mathrm{ref}\approx10$. Putting Eqs.~(\ref{eqn:A_Delta}) and (\ref{eqn:shaf_beta}) together, 
\begin{equation}
    A_\mathrm{ref}^2\stackrel{!}{=}\left(\hat{\mathbf{D}}_0+\frac{\delta\!\hat{\mathbf{D}}}{\delta\!\beta}\beta_\Delta\right)^2,
\end{equation}
where the hats indicate transformed quantities by the matrix $\pmb{M}$, Eq.~(\ref{eqn:M_matrix}). The critical $\beta_\Delta$ then corresponds to the largest negative root of this equation. That is, for every $\varphi$ we may define
\begin{equation}
    \beta_\Delta(\varphi)=\min\left[0, \beta_\pm:\left(\frac{\delta\!\hat{\mathbf{D}}}{\delta\!\beta}\right)^2\beta_\pm^2+2\frac{\delta\!\hat{\mathbf{D}}}{\delta\!\beta}\cdot\hat{\mathbf{D}}_0\beta_\pm+\left(\hat{\mathbf{D}}_0^2-\frac{1}{4r_\mathrm{ref}^2}\right)=0\right],\label{eq:beta-Delta}
\end{equation}
and $\beta_\pm$ may be computed explicitly with the quadratic formula. The single scalar $\beta_\Delta$ is then  the minimum value of this function in the whole $\varphi$ domain. 

\bibliographystyle{jpp}

\bibliography{jpp-instructions}

\end{document}